\newcommand{\cprime}{\/{\mathsurround=0pt$'$}}
\newcommand*{\Ev}{\mathbf{E}}
\newcommand*{\sym}{\mathrm{sym}\,}
\newcommand*{\abs}[1]{\left|#1\right|}

\newcommand*{\inv}{\mathrm{inv}}
\newtheorem{proposition}{Proposition}
\newtheorem{theorem}{Theorem}

\newtheorem{remark}{Remark}
\let\phi=\varphi
\documentclass[10pt]{iopart}
\usepackage{iopams}
\usepackage[all]{xy}
\SelectTips{cm}{}

\renewcommand{\pacs}{2010 \emph{Mathematics Subject Classification.}\ }

\usepackage{color}

\begin{document}
\title[Nonlocal symmetries of Lax integrable equations]{Nonlocal symmetries of
  Lax integrable equations: a comparative study}

\author{H Baran $^1$, I S Krasil{\cprime}shchik$^2$\footnote{Corresponding
    author},  O I Morozov$^3$ and P Voj{\v{c}}{\'{a}}k$^1$} \address{$^1$
  Mathematical Institute, Silesian University in Opava, Na Rybn\'{\i}\v{c}ku
  1, 746 01 Opava, Czech Republic} \address{$^2$ Independent University of
  Moscow, B. Vlasevsky 11, 119002 Moscow, Russia \& Russian State University
  for the Humanities, Miusskaya sq. 6, Moscow, GSP-3, 125993, Russia}
\address{$^3$ Faculty of Applied Mathematics, AGH University of Science and
  Technology, Al. Mickiewicza 30, Krak\'ow 30-059, Poland}

\ead{Hynek.Baran@math.slu.cz, josephkra@gmail.com, morozov@agh.edu.pl,
  Petr.Vojcak@math.slu.cz}

\begin{abstract}
  We continue here the study of Lax integrable equations. We consider four
  three-dimensional equations: (1) the rdDym
  equation~$u_{ty} = u_x u_{xy} - u_y u_{xx}$, (2) the 3D Pavlov equation
  $u_{yy} = u_{tx} + u_y u_{xx} - u_x u_{xy}$; (3) the universal hierarchy
  equation~$u_{yy} = u_t u_{xy} - u_y u_{tx}$, and (4) the modified Veronese
  web equation~$u_{ty} = u_t u_{xy} - u_y u_{tx}$. For each equation, using
  the known Lax pairs and expanding the latter in formal series in spectral
  parameter, we construct two infinite-dimensional differential coverings
  \cite{VinKrasTrends} and give a full description of nonlocal symmetry
  algebras associated to these coverings. For all the four pairs of coverings,
  the obtained Lie algebras of symmetries manifest similar (but not the same)
  structures: the are (semi) direct sums of the Witt algebra, the algebra of
  vector fields on the line, and loop algebras; all of them contain a
  component of finite grading. We also discuss actions of recursion operators
  on shadows of nonlocal symmetries.
\end{abstract}

\vspace{2pc}

% \submitto{\jpa}
\maketitle

\section*{Introduction and notation}
\label{sec:introduction}

In~\cite{BKMV-2014} we began a systematic study of symmetry and integrability
properties of Lax integrable 3D equations, i.e., equations that admit a Lax
pair with non-removable parameter. All the 2D symmetry reductions of
\begin{itemize}
\item the rdDym equation $u_{ty} = u_x u_{xy} - u_y u_{xx}$;
\item the 3D Pavlov equation $u_{yy} = u_{tx} + u_y u_{xx} - u_x u_{xy}$;
\item the universal hierarchy equation $u_{yy} = u_t u_{xy} - u_y u_{tx}$;
\item the modified Veronese web equation $u_{ty} = u_t u_{xy} - u_y u_{tx}$,
\end{itemize}
were described. In~\cite{BKMV-2015}, we studied the behavior of the Lax
operators admitted by these equations under symmetry reductions and showed
that in a number of cases the 2D reductions (one of them is equivalent to the
Gibbons-Tsarev equation~\cite{GT}) inherit the Lax integrability property. We
also constructed infinite series of (nonlocal) conservation laws for these
reductions. Finally, in the recent paper~\cite{BKMV-2016} we used expansion of
the Lax pair for the rdDym equation in formal series of the spectral parameter
to construct two infinite-dimensional differential coverings over this
equation and gave a full description of nonlocal symmetries in this covering.
All these equations are linearly degenerate in the sense of \cite{Fer-Mos},
where such equations  were classified.

In the current paper, we apply the same techniques to describe the Lie algebra
structure of nonlocal symmetries for the rest three
equations. \Sref{sec:preliminaries} is just a brief introduction to the
terminology used below. In \Sref{sec:rddym-equation}, to make the exposition
self-contained, we briefly recall the results obtained
in~\cite{BKMV-2016}. \Sref{sec:3d-pavlov-equation} is devoted to the 3D Pavlov
equation. The results on the universal hierarchy equation are discussed in
\Sref{sec:univ-hier-equat}, while the symmetries of the modified Veronese web
equation are described in \Sref{sec:modif-veron-web}. In each case, recursion
operators and their action on the shadows of nonlocal symmetries are also
discussed. We also describe an B\"{a}cklund auto-transformation for the
modified Veronese web equation. Finally, in \Sref{sec:discussion} we sum up
the results obtained. We omit the proofs: the reader can find a detailed
exposition for the case of the rdDym equation in~\cite{BKMV-2016} and all the
other proofs are quite similar.

All the symmetry algebras below have similar (but not the same) structure and
are direct or semi-direct sums of the following Lie algebras (see
\Tref{tab:all-the-algebras} on p.~\pageref{tab:all-the-algebras}, where the
main results are aggregated):
\begin{itemize}
\item the Witt algebra~$\mathfrak{W}$ of vector
  fields~$\mathbf{e}_i=z^{i+1}\frac{\partial}{\partial z}$,
   $i \in \mathbb{Z}$;
\item its subalgebras~$\mathfrak{W}_k^-$ spanned by~$\mathbf{e}_i$ with~$i\leq
  k\leq 0$ and~$\mathfrak{W}_k^+$ spanned by~$\mathbf{e}_i$ with~$i\geq
  k\geq 0$;
\item the algebra~$\mathfrak{V}[\rho]$ of vector
  fields~$R(\rho){\partial}/{\partial \rho}$ on~$\mathbb{R}^1$ with a
  distinguished coordinate~$\rho$; everywhere below we use the notation
  $[R,\bar{R}] = R\bar{R}' - \bar{R}R'$ for functions $R$ and
  $\bar{R}$ in $\rho$, where `prime' denotes the $\rho$-derivative;
\item the loop algebra~$\mathfrak{L}[\rho]$ spanned by the
  elements~$z^i\otimes X$, $i\in \mathbb{Z}$, $X\in \mathfrak{V}[\rho]$, with
  the commutator~$[z^i\otimes X, z^j\otimes Y]=z^{i+j}\otimes[X,Y]$;
\item the algebra~$\mathfrak{L}_k^+[\rho]$ spanned by the
  elements~$p(z)\otimes X$, where~$X\in \mathfrak{V}[\rho]$ and
  $p(z) \in \mathbb{R}[z]/(z^k)$ is a truncated polynomial. In a similar way,
  we define~$\mathfrak{L}_k^-[\rho]$ with~$p(z) \in \mathbb{R}[z^{-1}]/(z^{-k})$.
\end{itemize}
Semi-direct sums in the algebras of symmetries arise due to the natural
actions of~$\mathfrak{W}$ on~$\mathfrak{L}[\rho]$, $\mathfrak{W}_k^-$
on~$\mathfrak{L}_k^-[\rho]$, and~$\mathfrak{W}_k^+$
on~$\mathfrak{L}_k^+[\rho]$.

All the equations under consideration admit scaling symmetries that allow to
introduce natural weights (grading) to the space of polynomial functions on
the equation. This graded structure is inherited by the symmetry algebras in
all cases except for the modified Veronese web equation. Perhaps, this is the
reason why the Lie algebra structure of symmetries for this equation is a bit
different from the other ones.

\section{Preliminaries}
\label{sec:preliminaries}

Everywhere below we deal with second order scalar differential equations in
three independent variables~$x$, $y$, and~$t$. For a general coordinate-free
exposition see~\cite{AMS}. To this end, we consider the
space~$J^\infty(\mathbb{R}^3,\mathbb{R})$ of infinite jets of smooth
functions~$u=u(x,y,t)$ on~$\mathbb{R}^3$. This space is endowed with the
coordinates
\begin{equation*}
  x,\ y,\ t,\ u_{i,j,k}=\frac{\partial^{i+j+k}u}{\partial x^i\partial
    y^j\partial t^k},\quad i,j,k\geq 0,
\end{equation*}
and its geometric structure is determined by the Cartan distribution spanned
by the total derivatives
\begin{eqnarray*}
  D_x=\frac{\partial}{\partial x} +
  \sum_{i,j,k\geq0}u_{i+1,j,k}\frac{\partial}{\partial u_{i,j,k}},
  \quad
  D_y=\frac{\partial}{\partial y} +
  \sum_{i,j,k\geq0}u_{i,j+1,k}\frac{\partial}{\partial u_{i,j,k}},\\
  D_t=\frac{\partial}{\partial t} +
  \sum_{i,j,k\geq0}u_{i,j,k+1}\frac{\partial}{\partial u_{i,j,k}}.
\end{eqnarray*}
An equation~$\mathcal{E}=\{F=0\}\subset J^\infty(\mathbb{R}^3)$ is the subset
defined by the infinite system of relations $D_\sigma(F)=0$,
where~$F=F(x,y,t,u,u_x,u_y,u_t,u_{xx},u_{xy},\dots,u_{tt})$ is a smooth
function and~$D_\sigma$ denotes all possible compositions of the total
derivatives. Total derivatives and any differential operators in total
derivatives can be restricted to~$\mathcal{E}$, i.e., expressed in terms of
internal coordinates on~$\mathcal{E}$.

A symmetry of~$\mathcal{E}$ is a vector field
\begin{equation*}
  S=\sum S_{i,j,k}\frac{\partial}{\partial u_{i,j,k}}
\end{equation*}
on~$\mathcal{E}$ that commutes with the total derivatives (here and below
summation is taken over all internal coordinates on~$\mathcal{E}$). Any
symmetry is an evolutionary vector field of the form
\begin{equation*}
  \Ev_\phi=\sum D_x^iD_y^jD_t^k(\phi)\frac{\partial}{\partial u_{i,j,k}},
\end{equation*}
where~$\phi$ is an arbitrary smooth function on~$\mathcal{E}$ that satisfy the
equation $\ell_{\mathcal{E}}(\phi)=0$ and~$\ell_{\mathcal{E}}$ is the
restriction of the linearization operator
\begin{equation*}\fl
  \ell_F=\frac{\partial F}{\partial u} + \frac{\partial F}{\partial u_x}D_x +
  \dots + \frac{\partial F}{\partial u_t}D_t + \frac{\partial F}{\partial
    u_{xx}} D_x^2 +\frac{\partial F}{\partial u_{xy}}D_xD_y + \dots +
  \frac{\partial F}{\partial u_{tt}}D_t^2
\end{equation*}
to~$\mathcal{E}$. The function~$\phi$ is the generating function (or the
characteristic) of a symmetry.  Symmetries form a Lie
algebra~$\sym(\mathcal{E})$ with respect to commutator and the commutator
induces the Jacobi bracket on the space of generating functions:
$\{\phi_1,\phi_2\} = \Ev_{\phi_1}(\phi_2) - \Ev_{\phi_2}(\phi_s)$.  In what
follows, we do not distinguish between symmetries and their generating
functions.

A symmetry of the form
$s=\delta u + \alpha x u_x + \beta y u_y + \gamma t u_t$, $\alpha$, $\beta$,
$\gamma$, $\delta\in\mathbb{Z}$, is called a scaling symmetry
of~$\mathcal{E}$. If an equation admits such a symmetry one can introduce
weights to polynomial functions on~$\mathcal{E}$
by~$\abs{x}=-\alpha,\quad \abs{y}=-\beta$, $\abs{t}=-\gamma$,
$\abs{u_{i,j,k}}=\delta - i\alpha - j\beta-k\gamma$, with respect to which the
space~$\mathcal{P}(\mathcal{E})$ of such functions becomes graded:
$\mathcal{P}(\mathcal{E})=\oplus_{r\in\mathbb{Z}} \mathcal{P}_r(\mathcal{E})$.
If~$\Ev_\phi$ is a symmetry and~$\phi\in\mathcal{P}(\mathcal{E})$ we set
$\abs{\Ev_\phi}=\abs{\phi}-\abs{u}$.  Then
$\Ev_\phi(\mathcal{P}_r(\mathcal{E})) \subset
\mathcal{P}_{r+\abs{\Ev_\phi}}(\mathcal{E})$
and
$\big\vert[\Ev_{\phi_1},\Ev_{\phi_2}]\big\vert=\abs{\Ev_{\phi_1}} +
\abs{\Ev_{\phi_2}}$
and thus the space of polynomial symmetries becomes a $\mathbb{Z}$-graded Lie
algebra.

Let~$\mathcal{E}$ be an equation. A differential covering over~$\mathcal{E}$
(see~\cite{VinKrasTrends}) is an extension~$\tilde{\mathcal{E}}$
of~$\mathcal{E}$ by a system of first order equations
\begin{eqnarray}\nonumber
  w_x^\alpha&=X^\alpha(x,y,t,\dots,u_{i,j,k},\dots,w^\beta,\dots),\\\label{eq:26}
  w_y^\alpha&=Y^\alpha(x,y,t,\dots,u_{i,j,k},\dots,w^\beta,\dots),\\\nonumber
  w_t^\alpha&=T^\alpha(x,y,t,\dots,u_{i,j,k},\dots,w^\beta,\dots),
\end{eqnarray}
$\alpha$, $\beta=1,2,\dots$, such that this system is compatible
modulo~$\mathcal{E}$. The variables~$w^j$ are called nonlocal ones and there
exists a projection~$\tau\colon\tilde{\mathcal{E}} \to \mathcal{E}$ such that
the nonlocal variables are fiber-wise coordinates of this projection. The
number of independent nonlocal variables is the covering dimension. The total
derivatives are lifted to~$\tilde{\mathcal{E}}$ by
\begin{equation*}\fl
  \tilde{D}_x=D_x+\sum X^j\frac{\partial}{\partial w^j},\quad
  \tilde{D}_y=D_y+\sum Y^j\frac{\partial}{\partial w^j},\quad
  \tilde{D}_t=D_t+\sum T^j\frac{\partial}{\partial w^j}\quad
\end{equation*}
and consequently any differential operator~$D$ in total derivatives can be
lifted to~$\tilde{D}$ as well. We say that a covering is Abelian if the
right-hand sides of its defining equation do not depend on nonlocal variables.
In the case when system (\ref{eq:26}) may be written in the form of two
equations it is referred to as a Lax pair.

Given a one-dimensional covering~$\tau$ (i.e., a covering~\eref{eq:26}
with~$w^\alpha =w$, $X^\alpha=X$, $Y^\alpha=Y$, and~$T^\alpha=T$) that
smoothly depends on~$\lambda\in\mathbb{R}$, one can consider the expansion
$w=\sum_{-\infty}^\infty\lambda^i w_i$ and also expand the defining equations
of the covering in formal series of the parameter. Then an
infinite-dimensional covering with the nonlocal variables~$w_i$
arises. If~$w_i=0$ for~$i<0$ we say this is the positive covering associated
to~$\tau$; if~$w_i=0$ for~$i>0$ then we have the negative covering.

A symmetry of~$\tilde{\mathcal{E}}$ is a nonlocal symmetry
of~$\mathcal{E}$. Nonlocal symmetries are vector fields
\begin{equation*}
  \Ev_\phi + \sum_j \Phi^j\frac{\partial}{\partial w^j},
\end{equation*}
where~$\phi$, $\Phi^j$ are smooth functions on~$\tilde{\mathcal{E}}$ that
satisfy~$\tilde{\ell}_{\mathcal{E}}(\phi)=0$ together with the system
\begin{eqnarray*}\fl
  \tilde{D}_x(\phi^\alpha)=\tilde{\ell}_{X^\alpha}(\phi) + \sum_\theta
                            \frac{\partial
                            X^\alpha}{\partial w^j}\Phi^j,\qquad
  \tilde{D}_y(\phi^\alpha)=\tilde{\ell}_{Y^\alpha}(\phi) + \sum_\theta
                            \frac{\partial
                    Y^\alpha}{\partial w^j}\Phi^j,\\\fl
  \tilde{D}_t(\phi^\alpha)=\tilde{\ell}_{T^\alpha}(\phi) + \sum_\theta
                            \frac{\partial
                    T^\alpha}{\partial w^j}\Phi^j.
\end{eqnarray*}
A nonlocal symmetry is called invisible if~$\phi=0$.  Solutions of the
equation $\tilde{\ell}_{\mathcal{E}}(\phi)=0$ are called shadows.  We say that
a shadow~$\phi$ is lifted (or reconstructed) if there exists a nonlocal
symmetry~$\Phi=(\phi,\Phi^1,\dots,\Phi^j,\dots)$. Of course, lifts (if they
exist) are defined up to an invisible symmetry.

Let~$\mathcal{E}_1$, $\mathcal{E}_2$ be equations and~$\tau_i\colon
\tilde{\mathcal{E}} \to \mathcal{E}_i$ be coverings. Then one has the diagram
\begin{equation*}\xymatrixrowsep{1.4pc}
  \xymatrix{
    &\tilde{\mathcal{E}}\ar[ld]_-{\tau_1}\ar[rd]^-{\tau_2}&\\
    \mathcal{E}_1&&\mathcal{E}_2
  }
\end{equation*}
which is called a B\"{a}cklund transformation between~$\mathcal{E}_1$
and~$\mathcal{E}_2$. If~$\mathcal{E}_1=\mathcal{E}_2$ then it is an
B\"{a}cklund auto-transformation. To any equation~$\mathcal{E}=\mathcal{E}_F$,
we associate the system~$\mathcal{T}\mathcal{E}$
\begin{eqnarray*}
  &\fl F(x,y,y,u,u_x,u_y,u_t,u_{xx},u_{xy},\dots,u_{tt})=0,\\
  &\fl \ell_F(v)\equiv\frac{\partial F}{\partial u}v + \frac{\partial F}{\partial
    u_x}v_x + \frac{\partial F}{\partial u_y}v_y +\frac{\partial F}{\partial
    u_t}v_t + \frac{\partial F}{\partial u_{xx}}v_{xx} + \frac{\partial
    F}{\partial u_{xy}}v_{xy}  + \dots +  \frac{\partial F}{\partial
    u_{tt}}v_{tt} = 0,
\end{eqnarray*}
which is called the tangent equation of~$\mathcal{E}$. A B\"{a}cklund
auto-transformation of~$\mathcal{T}\mathcal{E}$ is a recursion operator for shadows
of symmetries of~$\mathcal{E}$; see~\cite{Mar-another}.

\section{The rdDym equation: a synopsis}
\label{sec:rddym-equation}

See~\cite{Blaszak,Ovsienko2010,Pavlov2006} for more information about the
equation. A detailed discussion of coverings, nonlocal symmetries, and
recursion operators for this equation can be found
in~\cite{BKMV-2016}. Neverthless, for the sake of completeness, we present a
short overview of the results obtained earlier.

The equation reads
\begin{equation}
  \label{eq:1}
  u_{ty}=u_xu_{xy}-u_yu_{xx}.
\end{equation}
We assign the weights $\abs{x} = 1$, $\abs{u} = 2$, $\abs{y} = \abs{t} = 0$ to
the variables $x$, $y$, $t$, and $u$; consequently the equation becomes
homogeneous with respect to these weights. Local symmetries are solutions of
the equation
$\ell_{\mathcal{E}}(\phi)\equiv D_tD_y(\phi) - u_xD_xD_y(\phi) +
u_yD_x^2(\phi) - u_{xy}D_x(\phi) + u_{xx}D_y(\phi) = 0$.
The space of solutions~$\sym(\mathcal{E})$ is spanned by the functions
\begin{eqnarray*}
  \psi_0= -x\,u_x + 2\,u,\qquad
  \upsilon_0(Y)=Yu_y,\\
  \theta_{0}(T)=Tu_t + T'(xu_x-u) + \frac{1}{2}T''x^2,\\
  \theta_{-1}(T)=Tu_x+T'x,\qquad
  \theta_{-2}(T)=T,
\end{eqnarray*}
where~$T=T(t)$, $Y=Y(y)$ are arbitrary functions of their arguments and
`prime' denotes the corresponding derivative. Commutators
of symmetries are presented in \Tref{tab:rdDym:loc-symm-comm}.
\begin{table}
  \caption{\label{tab:rdDym:loc-symm-comm}The rdDym equation: commutators of
    local symmetries.}
\begin{indented}
\item[]\begin{tabular}{@{}l|c|c|c|c|c}
         \br
         \rule{0em}{10pt}
         &$\psi_0$
         &$\upsilon_0(\bar{Y})$&$\theta_{0}(\bar{T})$&$\theta_{-1}(\bar{T})$
         &$\theta_{-2}(\bar{T})$\\\hline
         \rule{0em}{12pt}$\psi_0$
         &$0$&$0$&$0$&$
         \theta_{-1}(\bar{T})
         $&
         $
         2\theta_{-2}(\bar{T})
         $\\\hline
         \rule{0em}{12pt}$\upsilon_0(Y)$
         &\dots&$\upsilon_0([Y,\bar{Y}]) $&0&$0$&$0$\\\hline
         \rule{0em}{12pt}$\theta_{0}(T)$
         &\dots&\dots&$\theta_{0}([T,\bar{T}])$&$\theta_{-1}([\bar{T},T])$
         &$\theta_{-2}([\bar{T},T])$\\\hline
         \rule{0em}{12pt}$\theta_{-1}(T)$
         &\dots&\dots&\dots&$\theta_{-2}([\bar{T},T])$&$0$\\\hline
         \rule{0em}{12pt}$\theta_{-2}(T)$&\dots&\dots&\dots&\dots&$0$\\
         \br
       \end{tabular}
     \end{indented}
\end{table}
The corresponding evolutionary vector fields have the weights
$\abs{\Ev_{\psi_0}}=\abs{\Ev_{\upsilon_0(Y)}}=0$,
$\abs{\Ev_{\theta_{i}(T)}}=i$, $i=0,-1,-2$.

The system
\begin{equation}
  \label{eq:9}
  w_t=(u_x-\lambda)w_x\qquad
  w_y=\lambda^{-1}u_yw_x,
\end{equation}
is a Lax pair for \Eref{eq:1}. Setting
$w=\sum_{i=-\infty}^{+\infty}\lambda^iw_i$ and inserting this expansion
into~\eref{eq:9}, we obtain $w_{i,t}=u_xw_{i,x} - w_{i-1,x}$ and
$w_{i,y}=u_yw_{i+1,x}$.  The corresponding positive covering is defined by the
system
\begin{equation*}
  \begin{array}{lclclcl}
    q_{1,t}&=&\frac{u_x}{u_y},&&\quad q_{1,x}&=&\frac{1}{u_y},\\[4mm]
    q_{i,t}&=&\frac{u_x}{u_y}q_{i-1,y} - q_{i-1,x},
                              &&\quad
                                 q_{i,x}&=&\frac{q_{i-1,y}}{u_y},
  \end{array}
\end{equation*}
where $i\geq 2$, with the additional nonlocal variables~$q_i^{(j)}$ defined by
the equalities $q_i^{(0)}=q_i$ and $q_i^{(j+1)} = \left(q_i^{(j)}\right)_y$.
The weights assigned to the nonlocal variables are~${q_i^{(j)}} = -i$,
$i\geq 1$, $j\geq 0$.  The negative covering is defined by the system
\begin{equation*}
  \begin{array}{lclclcl}
    r_{1,x}&=&u_x^2-u_t,&&r_{1,y}&=&u_xu_y,\\[4mm]
    r_{i,x}&=&u_xr_{i-1,x}- r_{i-1,t},&&
    r_{i,y}&=&u_yr_{i-1,x}
  \end{array}
\end{equation*}
enriched by additional nonlocal variables~$r_i^{(j)}$ defined in the obvious
way by $r_i^{(0)}=r_i$, $r_i^{(j+1)}=\left(r_i^{(j)}\right)_t$.  One
has $\abs{r_i^{(j)}} = i+2$, $i\geq1$, $j\geq0$.

All the local symmetries of the rdDym equation can be lifted both to~$\tau^+$
and to~$\tau^-$ and we denote the lifts by the corresponding capital letters:
$\Psi_0$ for the lift of~$\psi_0$, $\Theta_i(T)$ for $\theta_i(T)$, etc.

Three families of nonlocal symmetries are admitted in~$\tau^+$. The first one
consists of invisible symmetries
\begin{equation*}
  \Phi_\inv^k(Y)= (\underbrace{0,\dots,0}_{k \textup{
      times}},\phi_\inv^1,\dots,\phi_\inv^i,\dots)
\end{equation*}
where~$\phi_\inv^1=Y(y)$, and another two are generated by the lifts
$\Psi_{-1}$ and $\Psi_{-2}$ of the nonlocal shadows $\psi_{-1}= q_1 u_y + x$
and $\psi_{-2}= (2q_2 - q_1 q_1^{(1)})u_y$ using the relations
$\Psi_{-k}=[\Psi_{-k+1},\Psi_{-1}]$, $k\geq3$, and
$\Upsilon_{-k}(Y)=[\Psi_{-k-1},\Phi_\inv^1(Y)]$.  The constructed nonlocal
symmetries have the weights $\abs{\Psi_i}=\abs{\Upsilon_{-l}(Y)}=i$,
$i\leq 0$, $\abs{\Theta_{j}(T)}=j$, $j=0$, $-1$, $-2$,
$\abs{\Phi_\inv^k(Y)}=k$, $k\geq 1$.

Then the following result is valid:
\begin{theorem}
  \label{thm:rdDym+}
  There exist a basis in $\sym_{\tau^+}(\mathcal{E})$ consisting of the
  elements $\{\mathbf{w}_i,\mathbf{v}_j(T),\mathbf{v}_k(Y)\}$\textup{,}
  $i\leq0$\textup{,} $j=0$\textup{,} $-1$\textup{,} $-2$,
  $k\in\mathbb{Z}$\textup{,} such that they commute as it is indicated in
  \Tref{tab:rdDym-commutators-tau+}.  So\textup{,} the
  algebra~$\sym_{\tau^+}(\mathcal{E})$ is isomorphic to
  $\mathfrak{W}_0^-\ltimes(\mathfrak{L}_3^-[t]\oplus\mathfrak{L}[y])$ with the
  natural action of~$\mathfrak{W}_0^-$
  on~$\mathfrak{L}_3^-[t]\oplus\mathfrak{L}[y]$.
\end{theorem}
\begin{table}
  \caption{\label{tab:rdDym-commutators-tau+}The rdDym equation: commutators
    in $\sym_{\tau^+}(\mathcal{E})$.}
  \begin{indented}
  \item[]\begin{tabular}{@{}l|c|c|c}
           \br
           \rule{0em}{10pt}
           &$\mathbf{w}_j$
           &$\mathbf{v}_j(\bar{T})$&$\mathbf{v}_j(\bar{Y})$\\\hline
           \rule{0em}{18pt}$\mathbf{w}_i$
           &$(j-i)\mathbf{w}_{i+j}$
           &$\begin{array}{ll}j\mathbf{v}_{i+j}(\bar{T}),&-2\leq i+j\leq0,\\
               0,&\textup{otherwise}\end{array}$
           &$j\mathbf{v}_{i+j}(\bar{Y})$\\\hline
           \rule{0em}{18pt}$\mathbf{v}_i(T)$
           &\dots
           &$\begin{array}{ll}\mathbf{v}_{i+j}([T,\bar{T}]),&-2\leq i+j\leq0,\\
               0,&\textup{otherwise}\end{array}$
           &$0$\\\hline
           \rule{0em}{11pt}$\mathbf{v}_i(Y)$
           &\dots&\dots&$\mathbf{v}_{i+j}([Y,\bar{Y}])$\\
           \br
         \end{tabular}
       \end{indented}
     \end{table}

In a similar way, local symmetries are lifted to~$\tau^-$ and three families
of nonlocal symmetries arise in this covering. They are
$\Psi_k$, $k\geq 1$, $\Theta_i(T)$, $i \geq -2$, $\Phi_\inv^l$ and have the
following weights: $\abs{\Psi_k}=k$, $k\geq 0$, $\abs{\Phi_\inv^l}=-l-2$,
$l\geq1$, $\abs{\Theta_i(T)}=i$, $i\geq -3$, $\abs{\Upsilon_0(Y)} =0$.

The Lie algebra structure is then described by

\begin{theorem}
  \label{thm:rdDym-}
  There exist a basis in $\sym_{\tau^-}(\mathcal{E})$ consisting of the
  elements $\{\mathbf{w}_i,\mathbf{v}_j(T),\mathbf{v}(Y)\}$\textup{,}
  $i\geq0$\textup{,} $j\in\mathbb{Z}$\textup{,} that satisfy the commutator
  relations presented in \Tref{tab:rdDym-comm-tau-}.  Hence\textup{,} the Lie
  algebra~$\sym_{\tau^-}(\mathcal{E})$ is isomorphic to
  $\mathfrak{W}_0^+\ltimes\mathfrak{L}[t]\oplus\mathfrak{V}[y]$ with the
  natural action of~$\mathfrak{W}_0^+$ on~$\mathfrak{L}[t]$.
\end{theorem}
\begin{table}
  \caption{\label{tab:rdDym-comm-tau-}The rdDym equation: commutators in
    $\sym_{\tau^-}(\mathcal{E})$.}
  \begin{indented}
  \item[]\begin{tabular}{@{}l|c|c|c}
           \br
           \rule{0em}{11pt}
           &$\mathbf{w}_j$&$\mathbf{v}_j(\bar{T})$&$\mathbf{v}(\bar{Y})$\\\hline
           \rule{0em}{11pt}$\mathbf{w}_i$
           &$(j-i)\mathbf{w}_{i+j}$&$j\mathbf{v}_{i+j}(\bar{T})$&$0$\\\hline
           \rule{0em}{11pt}$\mathbf{v}_i(T)$
           &\dots&$\mathbf{v}_{i+j}([T,\bar{T}])$&$0$\\\hline
           \rule{0em}{11pt}$\mathbf{v}(Y)$
           &\dots&\dots&$\mathbf{v}([Y,\bar{Y}])$\\
           \br
         \end{tabular}
       \end{indented}
     \end{table}

\begin{remark}
  Note that the components of the invisible symmetries are constructed using
  the operator
  \begin{equation*}
    \mathcal{Y}=q_1\frac{\partial}{\partial y} + \sum_{i=1}^\infty
  (i+1)q_{i+1}\frac{\partial }{\partial q_i}.
  \end{equation*}
  Similar operators will arise in the study of other equations below.
\end{remark}

The algebra~$\sym(\mathcal{E})$ admits a recursion
operator~$\hat{\chi} = \mathcal{R}_{+}(\chi)$ defined by the system
\begin{equation}\label{recursion_operator}
  \begin{array}{rcl}
    D_t (\hat{\chi}) &=&
                         u_y^{-1}\,\big(u_y\,D_x(\chi)-
                         u_x\,D_y(\chi)+(u_xu_{xy}-u_yu_{xx}) \hat{\chi}\big),
                             \\[2mm]
    D_x (\hat{\chi})&=& u_y^{-1}\,\big(u_{xy}\,\hat{\chi}-D_y(\chi)\big),
  \end{array}
\end{equation}
see~\cite{Morozov2012}.  This means that~$\hat{\chi}$ is a nonlocal shadow
whenever~$\chi$ is.  Another recursion
operator~$\chi= \mathcal{R}_{-}(\hat{\chi})$ is given by the system
\begin{equation} \label{inverse_recursion_operator}
  \begin{array}{rcl}
    D_x (\chi) &=& D_t(\hat{\chi}) - u_x\,D_x(\hat{\chi}) +u_{xx}\,\hat{\chi},
    \\[2mm]
    D_y (\chi) &=& - u_y\,D_x(\hat{\chi})+ u_{xy}\,\hat{\chi}.
  \end{array}
\end{equation}
The operators~$\mathcal{R}_+$ and~$\mathcal{R}_-$ are mutually inverse.

The actions of $\mathcal{R}_{+}$ and $\mathcal{R}_{-}$ on $\sym(\mathcal{E})$
may be prolonged to the shadows of nonlocal symmetries from
$\sym(\tilde{\mathcal{E}}^+)$ and $\sym(\tilde{\mathcal{E}}^-)$ if we replace
the derivatives $D_t$, $D_x$ and $D_y$ in~\eref{recursion_operator}
and~\eref{inverse_recursion_operator} by the total terivatives $\hat{D}_t$,
$\hat{D}_x$ and $\hat{D}_y$ in the Whitney product of the coverings $\tau^+$
and $\tau^-$ in the sense of~\cite{VinKrasTrends}. The resulting operators
will be also denoted by $\mathcal{R}_{+}$ and $\mathcal{R}_{-}$.

Note that the operators act nontrivially on `vacuum':
$\mathcal{R}_+(0)=\theta_{-2}(T)$, $\mathcal{R}_-(0)=\upsilon_0(Y)$, which
immediately follows from Equations~\eref{recursion_operator}
and~\eref{inverse_recursion_operator}; thus the actions are reasonable to
consider modulo~$\theta_{-2}(T)$ for~$\mathcal{R}_+$ and~$\upsilon_0(Y)$
for~$\mathcal{R}_-$. Taking into account this remark, we have the following

\begin{proposition}
  Modulo the images of the trivial symmetry\textup{,} the action of recursion
  operators is of the form
  \begin{eqnarray*}
    &\fl\mathcal{R}_+(\theta_i(T))=\left\{\begin{array}{ll}
      \alpha_i^+\theta_{i-1}(T),
      &i>-2,\\
      0,&i=-2,
    \end{array}\right.
      &\quad\mathcal{R}_-(\theta_i(T))=\alpha_i^-\theta_{i+1}(T),\quad
         i\geq-2,
    \\
    &\fl\mathcal{R}_+(\upsilon_i(Y))=\beta_i^+\upsilon_{i+1}(Y),\quad i\leq 0,
      &\quad\mathcal{R}_-(\upsilon_i(Y))=\left\{
         \begin{array}{ll}
           \beta_i^-\upsilon_{i+1}(Y),&i<0,\\
           0,&i=0,
         \end{array}\right.
    \\
    &\fl\mathcal{R}_+(\psi_i)=\gamma_i^+\psi_{i-1},
      &\quad\mathcal{R}_-(\psi_i)=\gamma_i^-\psi_{i+1},
         \qquad i\in \mathbb{Z},
  \end{eqnarray*}
  where~$\alpha_i^\pm$\textup{,} $\beta_i^\pm$\textup{,} and~$\gamma_i^\pm$
  are nonzero constants.
\end{proposition}

Note that the recursion operators $\mathcal{R}_{+}$ and $\mathcal{R}_{-}$
`glue together' the shadows $\psi_m$ of nonlocal symmetries in coverings
$\tilde{\mathcal{E}}^{+}$ and $\tilde{\mathcal{E}}^{-}$ and `tunnel' from the
series of $\theta_k(T)$ to that of $\upsilon_{j}(Y)$, see \Fref{tab:RO-act}.

\begin{remark}\label{sec:rddym-equat-synops-rem}
  In all the figures here and below straight arrows denote actions up to
  scalar multipliers and modulo the image of the trivial shadow. We also
  `compress' the notation and write $\theta_i$ instead of
  $\theta_i(T)$\textup{,} $\upsilon_k$ instead of $\upsilon_k(Y)$\textup{,}
  etc. Notation $(\cdot)^+$ means that a shadow lives in $\tau^+$, $(\cdot)^-$
  is for those who live in $\tau^-$; shadows marked by $(\cdot)^\pm$ live in
  both coverings.
\end{remark}

\begin{figure}
  \centering
  \begin{equation*}\fl\xymatrixrowsep{1pc}
    \xymatrix{
      &\dots\ \ar@<1ex>[r]^-{\mathcal{R}_-}
      &\ar[l]^-{\mathcal{R}_+}\psi_{-1}^+\ar@<1ex>[r]^-{\mathcal{R}_-}
      &\ar[l]^-{\mathcal{R}_+}\psi_0^\pm\ar@<1ex>[r]^-{\mathcal{R}_-}
      &\ar[l]^-{\mathcal{R}_+}\psi_1^-\ar@<1ex>[r]^-{\mathcal{R}_-}
      &\ar[l]^-{\mathcal{R}_+}\ \dots&&\\
      \dots\ \ar@<1ex>[r]^-{\mathcal{R}_-}
      &\ar[l]^-{\mathcal{R}_+}\upsilon_{-1}^+\ar@<1ex>[r]^-{\mathcal{R}_-}
      &\ar[l]^-{\mathcal{R}_+}\upsilon_0^\pm\ar@<1ex>[r]^-{\mathcal{R}_-}
      &\ar[l]^-{\mathcal{R}_+}0^\pm\ar@<1ex>[r]^-{\mathcal{R}_-}
      &\ar[l]^-{\mathcal{R}_+}\theta_{-2}^\pm\ar@<1ex>[r]^-{\mathcal{R}_-}
      &\ar[l]^-{\mathcal{R}_+}\theta_{-1}^\pm\ar@<1ex>[r]^-{\mathcal{R}_-}
      &\ar[l]^-{\mathcal{R}_+}\theta_0^\pm\ar@<1ex>[r]^-{\mathcal{R}_-}
      &\ar[l]^-{\mathcal{R}_+}\theta_1^-\ar@<1ex>[r]^-{\mathcal{R}_-}
      &\ar[l]^-{\mathcal{R}_+}\dots
    }
  \end{equation*}
  \caption{The rdDym equation: action of recursion operators
    \eref{recursion_operator} and \eref{inverse_recursion_operator}}
  \label{tab:RO-act}
\end{figure}

\section{The 3D Pavlov equation}
\label{sec:3d-pavlov-equation}

This equation was discussed in~\cite{Dun,Pav}, for example.

\subsection{The equation}
\label{sec:equation-1}

The 3D Pavlov equation is of the form
\begin{equation}
  \label{eq:2}
  u_{yy}=u_{tx}+u_yu_{xx}-u_xu_{xy}.
\end{equation}
We choose the following internal coordinates on~$\mathcal{E}$:
\begin{equation*}
  u_{k,l}^0=u_{\underbrace{x\dots x}_{k\textup{\tiny\ times}}\underbrace{t\dots
      t}_{l\textup{\tiny\ times}}},\quad u_{k,l}^1=u_{\underbrace{x\dots
      x}_{k\textup{\tiny\
        times}}
    \underbrace{t\dots
      t}_{l\textup{\tiny\ times}}},  \qquad k,l\geq 0.
\end{equation*}
Then the total derivatives read
\begin{eqnarray*}
  D_x&=\frac{\partial}{\partial x}
       +\sum_{k,l}\left(u_{k+1,l}^0\frac{\partial}{\partial u_{k,l}^0} +
       u_{k+1,l}^1\frac{\partial}{\partial u_{k,l}^1}\right),\\
  D_y&=\frac{\partial}{\partial y} +
       \sum_{k,l}\left(u_{k,l}^1\frac{\partial}{\partial u_{k,l}^0}
       +D_x^kD_t^l\left(u_{11}^0+u_{00}^1u_{20}^0-u_{10}^0u_{10}^1\right)
       \frac{\partial}{\partial
       u_{k,l}^1}\right),\\
  D_t&=\frac{\partial}{\partial
       t}+\sum_{k,l}\left(u_{k,l+1}^0\frac{\partial}{\partial u_{k,l}^0} +
       u_{k,l+1}^1\frac{\partial}{\partial u_{k,l}^1}\right)
\end{eqnarray*}
in these coordinates.  We assign the weights $\abs{t}=0$, $\abs{y}=1$,
$\abs{x}=2$, $\abs{u}=3$ and hence $\abs{u_{k,l}^0}=3-2k$,
$\abs{u_{k,l}^1} = 3-2k-1$.

Symmetries of~$\mathcal{E}$ are solutions to the equation
\begin{equation}
  \label{eq:6}\fl
  \ell_{\mathcal{E}}(\phi)\equiv
  D_y^2(\phi)-D_tD_x(\phi)-u_yD_x^2(\phi)+u_xD_xD_y(\phi)-u_{xx}D_y(\phi)
  +u_{xy} D_x(\phi).
\end{equation}
The space~$\sym(\mathcal{E})$ of solutions to \Eref{eq:6} is spanned by the
functions
\begin{eqnarray*}\fl
  \phi_1=2x-yu_x,\qquad \phi_2=3u-2xu_x-yu_y,\\\fl
  \theta_0(T)=Tu_t+T'(xu_x+yu_y-u)+\frac{1}{2}T''(y^2u_x-2xy)-
  \frac{1}{6}T'''y^3, \\\fl
  \theta_1(T)=Tu_y+T'(yu_x-x)-\frac{1}{2}T''y^2,\qquad
  \theta_2(T)=Tu_x-T'y,\qquad
  \theta_3(T)=T,
\end{eqnarray*}
where~$T$ is a function of~$t$ and `prime' denotes the
$t$-derivatives. Commutators of these symmetries are presented in
\Tref{tab:Pavlov-loc-sym-comm}.
\begin{table}
\caption{\label{tab:Pavlov-loc-sym-comm}The Pavlov equation: commutators of
  local symmetries.}
\begin{indented}
\item[]\begin{tabular}{@{}l|c|c|c|c|c|c}
         \br
         \rule{0em}{11pt}
         &$\phi_1$
         &$\phi_2$
         &$\theta_0(\bar{T})$
         &$\theta_1(\bar{T})$&$\theta_2(\bar{T})$&$\theta_3(\bar{T})$
         \\\hline
         \rule{0em}{11pt}$\phi_1$
         &$0$&$\phi_1$&$0$&$-2\theta_2(\bar{T})$&$2\theta_3(\bar{T})$&$0$
         \\\hline
         \rule{0em}{11pt}$\phi_2$
         &\dots
         &$0$
         &$0$
         &$-\theta_1(\bar{T})$&$-2\theta_2(\bar{T})$&$-3\theta_3(\bar{T})$
         \\\hline
         \rule{0em}{11pt}$\theta_0(T)$
         &\dots
         &\dots
         &$\theta_0([\bar{T},T])$
         &$\theta_1([\bar{T},T])$
                             &$\theta_2([\bar{T},T])$&$\theta_3([\bar{T},T])$
         \\\hline
         \rule{0em}{11pt}$\theta_1(T)$
         &\dots
         &\dots&\dots&$\theta_2([\bar{T},T])$&$\theta_3([\bar{T},T])$&$0$
         \\\hline
         \rule{0em}{11pt}$\theta_2(T)$&\dots&\dots&\dots&$\dots$&$0$&$0$
         \\\hline
         \rule{0em}{11pt}$\theta_3(T)$&\dots&\dots&\dots&\dots&\dots&$0$\\
         \br
       \end{tabular}
     \end{indented}
   \end{table}
The corresponding vector fields have the weights~$\abs{\Ev_{\phi_1}}=-1$,
$\abs{\Ev_{\phi_2}}=0$, $\abs{\Ev_{\theta_i}}=-i$, $i=0,\dots, -3$.

\subsection{The Lax pair and hierarchies}
\label{sec:lax-pair-hierarchies-1}

The Lax pair for the 3D Pavlov equation is
$q_{t}=(\lambda^2-\lambda u_x-u_y)q_x$, $q_{y}=(\lambda-u_x)q_{x}$.
Expanding~$q$ in integer powers of~$\lambda$, we arrive to the covering
$q_{i,t}=q_{i-2,x}-u_xq_{i-1,x}-u_yq_{i,x}$, $q_{i,y}=q_{i-1,x}-u_xq_{i,x}$,
for all $i\in\mathbb{Z}$.

The positive covering corresponding to this system is
\begin{eqnarray*}
  &q_{0,t}+u_yq_{0,x} =0, &\ q_{0,y} +u_xq_{0,x}= 0;\\
  &q_{1,t}+u_yq_{1,x} =-u_xq_{0,x},&\ q_{1,y} +u_xq_{1,x}= q_{0,x};\\
  &q_{i,t}+u_yq_{i,x} =q_{i-2,x}-u_xq_{i-1,x}, &\ q_{i,y}+u_xq_{i,x}=q_{i-1,x},
\end{eqnarray*}
where $i\geq 2$, to which nonlocal variables~$q_i^{(j)}$ defined by
$q_i^{(0)}=q_i$, $q_i^{(j+1)}=q_{i,x}^{(j)}$ are added. One
has~$\abs{q_i^{(j)}} = -i - 2j$.  This covering is not Abelian.

The negative covering is given by
\begin{equation*}
  \begin{array}{lclclcl}
    r_{1,y}&=&u_t+u_xu_y,&&r_{1,x}&=&u_y+u_x^2;\\
    r_{i,y}&=&r_{i-1,t}+u_yr_{i-1,x},&&r_{i,x}&=&r_{i-1,y}+u_xr_{i-1,x},
  \end{array}
\end{equation*}
$i\geq 2$, with additional nonlocal variables~$r_i^{(j)}$ defined by
$r_i^{(0)}=r_i$, $r_i^{(j+1)}=r_{i,t}^{(j)}$.  One has~$\abs{r_i^{(j)}}=i+3$.

\subsection{Nonlocal symmetries in the positive covering}
\label{sec:nonl-symm-posit-1}

\subsubsection{Lifts of local symmetries}
\label{sec:lifts-local-symm-4}

All the local symmetries can be lifted to~$\tau^+$. In more detail, we have
the following results.

The lift of~$\phi_1=yu_x-2x$ is
$\Phi_1=(\phi_1,\phi_1^0,\dots,\phi_1^i,\dots)$,
where~$\phi_1^i = yq_{i,x} + (i + 1)q_{i+1}$.  The
symmetry~$\phi_2= 2xu_x + yu_y - 3u$ is lifted by
$\Phi_2=(\phi_2,\phi_2^0,\dots,\phi_2^i,\dots)$,
where~$\phi_2^0 = -\phi_1 q_{0,x}$
and~$\phi_2^i = -\phi_1 q_{i,x} + yq_{i-1,x} + iq_i$, $i\geq1$.  The lift
of~$\theta_2(T)=Tu_x-T'y$ is
$\Theta_2(T)=(\theta_2,Tq_{0,x},\dots,Tq_{i,x},\dots)$.  The symmetry
\begin{equation*}
  \theta_1(T) = Tu_y + T'(yu_x - x) - \frac{1}{2} T''y^2
\end{equation*}
admits the lift
$\Theta_1(T)=(\theta_1, \theta_1^0, \theta_1^1 ,\dots \theta_1^i ,\dots)$,
where~$\theta_1^0 = -\theta_1(T)q_{0,x}$
and~$\theta_1^i = -\theta_2(T)q_{i,x} + Tq_{i-1,x}$, $i\geq1$.  The lift of
\begin{equation*}
  \theta_0(T) = T u_t + T' (xu_x + yu_y - u) + T''\left(\frac{1}{2}y^2 u_x
    - xy\right)- \frac{1}{6}T''' y^3
\end{equation*}
is
$\Theta_0(T)=(\theta_0,\theta_0^0,\theta_0^1,\theta_0^2,\dots,\theta_0^i,\dots)$,
where $\theta_0^0 = -\theta_1(T)q_{0,x}$,
$\theta_0^1 = -\theta_1(T)q_{1,x} - \theta_2(T)q_{0,x}$, and
$\theta_0^i = -\theta_1(T)q_{i,x} - \theta_2(T)q_{i-1,x} + Tq_{i-2,x}$,
$i\geq2$.  Finally, for~$\theta_3(T)=T$ one
has~$\Theta_3(T)=(\theta_3,0\dots,0,\dots)$.

\subsubsection{Nonlocal symmetries}
\label{sec:nonlocal-symmetries-4}

Three families of nonlocal symmetries exist for the Pavlov equation
in~$\tau^+$. The first one consists of invisible symmetries
\begin{equation*}
  \Phi^k_\inv(Y) = (\underbrace{0,...,0}_{k\textup{\tiny\
      times}},\phi_\inv^{k,k+1},\phi_\inv^{k,k+2} ,\dots,
  \phi_\inv^{k,k+i},\dots),\qquad
  k = 1,2,\dots,
\end{equation*}
where for every $i \ge 1$ it holds~$\phi_\inv^{k,k+i}=R_{i-1}(Q)$.  Here
$R_0(Q)=Q(q_0)$ is an arbitrary function of $q_0$ and for $n \ge 1$ we define
\begin{equation*}
  R_n(Q) = \frac{1}{n}\,\mathcal{Y} (R_{n-1}(Q)),
\end{equation*}
where $\mathcal{Y}$ is the vector field
\begin{equation*}
  \mathcal{Y} =\sum_{i=0}^\infty(i + 1)q_{i+1}\frac{\partial}{\partial q_i}.
\end{equation*}
Now we define explicitly the nonlocal symmetry
$\Psi_{-1}=(\psi_1,\psi_1^0,\dots,\psi_1^i,\dots)$ by setting
\begin{equation*}
  \psi_{-1} = \frac{q_1}{q_{0,x}} + y
\end{equation*}
and
\begin{equation*}
  \psi_{-1}^i = -(i + 2)q_{i+2} + \frac{q_1q_{i+1,x}}{q_{0,x}}.
\end{equation*}
Then the elements of the second nonlocal family are
$\Psi_{-k}=[\Phi_1,\Psi_{-1}]$, $k\geq 2$.  One has $\abs{\Psi_{-k}} = -k-1$.
Finally, we define $\Xi_l(Q) = [\Psi_{-l},\Phi_\inv^2(Q)]$, $l\geq 1$.

Distribution of symmetries along weights is $\abs{\Psi_l}=-l-1,l\geq1$,
$\abs{\Phi_1}=-1$, $\abs{\Phi_2}=0$, $\abs{\Theta_k(T)}=k-2$, $k=0,\dots,3$,
$\abs{\Xi_j(Q)}=-j+1$, $j\geq1$, $\abs{\Phi_l^\inv(Q)}=l$, $l\geq 1$.

\subsubsection{Lie algebra structure}
\label{sec:lie-algebra-struct-4}

Consider the spaces $W$ spanned by $\Phi_1$, $\Phi_2$, and $\Psi_i$,$i\leq -1$,
$V[t]$ spanned by $\Theta_i(T)$, $i=0,\dots,3$, $V[q_0]$ spanned by
$\Phi_\inv^i(Q)$ and $\Xi_j(Q)$, $i$, $j\geq1$. Then the following result
holds:

\begin{theorem}
  \label{thm:Pavlov+}
  There exist bases $\mathbf{w}_i$ in $W$\textup{,} $i\leq 0$\textup{,}
  $\mathbf{v}_i(T)$ in $V[t]$\textup{,} $i=0,\dots,-3$\textup{,} and
  $\mathbf{v}_i(Q)$ in $V[q_0]$\textup{,} $i\in\mathbb{Z}$\textup{,} such that
  their commutators satisfy the relations presented in
  \Tref{tab:Pavlov-lie-alg-structure-+}.  In other words\textup{,}
  $\sym_{\tau^+}(\mathcal{E})$ is isomorphic to
  $\mathfrak{W}_0^-\ltimes (\mathfrak{L}[q_0]\oplus\mathfrak{L}_4^-[t])$ with
  the natural action of the Witt algebra $\mathfrak{W}_0^-$
  on~$\mathfrak{L}[q_0]\oplus\mathfrak{L}_4^-[t]$.
\end{theorem}
\begin{table}
  \caption{\label{tab:Pavlov-lie-alg-structure-+}The Pavlov equation:
    commutators in $\sym_{\tau^+}(\mathcal{E})$.}
\begin{indented}
\item[]\begin{tabular}{@{}l|c|c|c}
\br
    &$\mathbf{w}_j$&$\mathbf{v}_j(\bar{T})$&$\mathbf{v}_j(\bar{Q})$\\\hline
    \rule{0em}{17pt}$\mathbf{w}_i$
    &$(j-i)\mathbf{w}_{i+j}$
    &$\begin{array}{ll}j\mathbf{v}_{i+j}(\bar{T}),&-3\leq i+j\leq0\\
        0,&\textup{otherwise}\end{array}$
                   &$j\mathbf{v}_{i+j}(\bar{Q})$\\\hline
    \rule{0em}{17pt}$\mathbf{v}_i(T)$
    &\dots&$\begin{array}{ll}(j-i)\mathbf{v}_{i+j}([T,\bar{T}]),&-3\leq
                                                                  i+j\leq0\\
                                               0,&\textup{otherwise}
                                             \end{array}$
                   &$0$\\\hline
    \rule{0em}{11pt}$\mathbf{v}_i(Q)$
    &\dots&\dots&$\mathbf{v}_{i+j}([Q,\bar{Q}])$\\
\br
\end{tabular}
\end{indented}
\end{table}

\subsection{Nonlocal symmetries in the negative covering}
\label{sec:nonl-symm-negat-1}

\subsubsection{Lifts of local symmetries}
\label{sec:lifts-local-symm-5}

Similar to the case of~$\tau^+$, all the local symmetries are lifted to the
covering~$\tau^-$. Namely, the symmetry~$\phi_1=yu_x-2x$ has the lift
$\Phi_1 = (\phi_1, \phi_1^1 , \phi_1^2,\dots,\phi_1^i,\dots)$,
where~$\phi_1^1 = yr_{1,x} - 3u$ and~$\phi_1^i = yr_{i,x} - (i + 2)r_{i-1}$,
$i\geq2$.  The symmetry $\phi_2 = 2xu_x + yu_y - 3u$ has the lift
$\Phi_2 = (\phi_2, \phi_2^1, \phi_2^2,\dots,\phi_2^i,\dots)$,
where~$\phi_2^i = 2xr_{i,x} + yr_{i,y} - (i + 3)r_i$, $i\geq 1$.

To describe the lift
$\Theta_3(T)=(\theta_3,\theta_3^1,\theta_3^2,\dots,\theta_3^i,\dots)$
of~$\theta_3(T)=T$, consider the operator
\begin{equation}
  \label{eq:19}
  \mathcal{Y} = y\frac{\partial }{\partial t} + 2x\frac{\partial}{\partial y}
  + 3u\frac{\partial }{\partial x} + 4q_1\frac{\partial }{\partial u} +
  \sum_{i=1}^\infty (i + 4)q_{i+1}\frac{\partial }{\partial q_i}
\end{equation}
and set~$\theta_3^1=yT'$ and
$\theta_3^i=\frac{1}{i}\mathcal{Y}(\phi_6^{i-1})$, $i\geq 2$.

To describe the lifts of~$\theta_2(T)=Tu_x-T'y$,
$\theta_1(T)= Tu_y + T'(yu_x - x) - \frac{1}{2} T'' y^2$,
and~$\theta_0(T) = Tu_t + T'(xu_x + yu_y - u) + T''\left(\frac{1}{2}y^2 u_x -
  xy\right) - \frac{1}{6}T'''y^3 3$,
we shall need the nonlocal symmetry~$\Psi_0$ (see \Eref{eq:20} below). Namely,
we set
\begin{equation*}\fl
  \Theta_2(T) = \frac{1}{3}[\Psi_0,\Theta_3(T)],
     \quad
  \Theta_1(T) = - \frac{1}{2}  [\Psi_0,\Theta_2(T)],
     \quad
  \Theta_0(T) = -[\Psi_0,\Theta_1(T)].
\end{equation*}

\subsubsection{Nonlocal symmetries}
\label{sec:nonlocal-symmetries-5}

The invisible symmetries in~$\tau^-$ are of the form
\begin{equation*}
  \Phi_\inv^k(T)= (\underbrace{0,\dots,0}_{k
    \textup{\tiny\ times}},\phi_\inv^{k,k+1},\dots,\phi_\inv^{k,k+i},\dots),
\end{equation*}
where for every $i \ge 1$ it holds $\phi_\inv^{k,k+i} =R_{i-1}(T)$ and the
sequence of functions $R_n$, $n \ge 0$ is defined as follows
\begin{equation*}
R_0(T) = T,
\qquad
R_{n+1}(T) = \frac{1}{n+1}\,\mathcal{Y}(R_{n}(T))
\end{equation*}
with the operator~$\mathcal{Y}$ being defined by \Eref{eq:19}.

Let us now introduce the nonlocal symmetries
\begin{equation}
  \label{eq:20}
       \Psi_0=(\psi_0,\psi_0^1,\dots,\psi_0^i,\dots)
\end{equation}
and $\Psi_1=(\psi_1,\psi_1^1,\dots,\psi_1^i,\dots)$ by setting
$\psi_0= 4r_1 - 3uu_x - 2xu_y - yu_t$,
$\psi_1 = 5r_2 - 4u_xr_1 - yr_{1,t} - 3uu_y - 2xu_t + yu_tu_x$, and
$\psi_0^i = (i + 4)r_{i+1} - 3ur_{i,x} - 2xr_{i,y} - yr_{i,t}$,
$\psi_1^i= (i + 5)r_{i+2} - yr_{i+1,t} - 3ur_{i,y} - 2xr_{i,t} - (4r_1 -
yu_t)r_{i,x}$ for $i\geq 1$.

Using the symmetries~$\Psi_0$ and~$\Psi_1$, we define by induction two new
families of nonlocal symmetries by $\Psi_k=[\Psi_0,\Psi_{k-1}]$, $k\geq 2$,
and $\Omega_l(T)=[\Psi_l,\Theta_1(T)]$.

The weights of the obtained symmetries are $\abs{\Phi_1}=-1$,
$\abs{\Phi_2}=0$, $\abs{\Psi_k}=k+1$, $k\geq0$, $\abs{\Omega_l(T)}=l$, $l\geq1$,
$\abs{\Phi_l^\inv(T)}=-l-3$, $l\geq1$, $\abs{\Theta_i(T)}=-i$, $i=0,\dots, 3$.

\subsubsection{Lie algebra structure}
\label{sec:lie-algebra-struct-5}

Consider the subspaces $W$ spanned by $\Phi_1$, $\Phi_2$, $\Psi_i$, $i\geq0$,
and $V[t]$ spanned by $\Phi_\inv^i(T)$, $\Omega_j(T)$, $i$, $j\geq1$,
$\Theta_k(T)$, $k=0,\dots,2$ in~$\sym_{\tau^-}(\mathcal{E})$.

\begin{theorem}
  \label{thm:Pavlov-}
  There exist bases $\mathbf{w}_i$ in $W$\textup{,} $i\geq-1$\textup{,} and
  $\mathbf{v}_j(T)$ in $V[t]$\textup{,} $j\in\mathbb{Z}$\textup{,} that
  satisfy the commutator relations indicated in
  \Tref{tab:Pavlov-Lie-commutators-in-tau-}.  In other words\textup{,} the Lie
  algebra~$\sym_{\tau^-}(\mathcal{E})$ is isomorphic to
  $\mathfrak{W}_{-1}^+\ltimes \mathfrak{L}[t]$ with the natural action
  of~$\mathfrak{W}_{-1}^+$ on~$\mathfrak{L}[t]$.
\end{theorem}
\begin{table}
  \caption{\label{tab:Pavlov-Lie-commutators-in-tau-}The Pavlov equation:
    commutators in $\sym_{\tau^-}(\mathcal{E})$.}
\begin{indented}
\item[]\begin{tabular}{@{}l|c|c}
\br
    \rule{0em}{10pt}
    &$\mathbf{w}_j$&$\mathbf{v}_j(\bar{T})$\\\hline
    \rule{0em}{10pt}$\mathbf{w}_i$
    &$(j-i)\mathbf{w}_{i+j}$&$j\mathbf{v}_{i+j}(\bar{T})$\\\hline
    \rule{0em}{10pt}$\mathbf{v}_i(T)$
    &\dots&$\mathbf{v}_{i+j}([T,\bar{T}])$\\
\br
\end{tabular}
\end{indented}
\end{table}

\subsection{Recursion operators}
\label{sec:recursion-operators-1}
We have the following result (see~\cite{Morozov2012}):

\begin{proposition}
  \Eref{eq:2} admits the recursion operator for symmetries
  $\psi=\mathcal{R}_{+}(\varphi)$ defined by the following system\textup{:}
  \begin{equation}\fl
    D_t(\psi) = -u_y\,D_x(\psi)+u_{xy}\,\psi+D_y(\varphi),
    \quad
    D_y(\psi) = -u_x\,D_x(\psi)+u_{xx}\,\psi+D_x(\varphi).
    \label{direct_recursion_operator_for_Pavlov_eq}
  \end{equation}
  The inverse operator $\varphi=\mathcal{R}_{-}(\psi)$ is defined by system
  \begin{equation}\fl
    D_x(\varphi) = u_x\,D_x(\psi)+D_y(\psi)-u_{xx}\,\psi,
    \quad
    D_y(\varphi) = D_t(\psi)+u_y\,D_x(\psi)-u_{xy}\,\psi.
    \label{inverse_recursion_operator_for_Pavlov_eq}
  \end{equation}
\end{proposition}
The action of the recursion operators on shadows is schematically shown in
\Fref{tab:RO-act-Pavlov_eq}, see Remark \ref{sec:rddym-equat-synops-rem} for
notation. Here $\xi_i^{+}$ and $\omega_i^{-}$ are the shadows of $\Xi_i(T)$
and $\Omega_i(T)$, respectively.

\noindent\begin{figure}[h]
  \begin{equation*}
    \xymatrixcolsep{1.1pc}\xymatrixrowsep{1.5pc}\fl
    \xymatrix{
      &\dots\ \ar@<1ex>[r]^-{\mathcal{R}_-}
      &\ar[l]^-{\mathcal{R}_+}\psi_{-2}^+\ar@<1ex>[r]^-{\mathcal{R}_-}
      &\ar[l]^-{\mathcal{R}_+}\psi_{-1}^+\ar@<1ex>[r]^-{\mathcal{R}_-}
            &\ar[l]^-{\mathcal{R}_+}\phi_1^\pm\ar@<1ex>[r]^-{\mathcal{R}_-}
      &\ar[l]^-{\mathcal{R}_+}\phi_2^\pm\ar@<1ex>[r]^-{\mathcal{R}_-}
      &\ar[l]^-{\mathcal{R}_+}\psi_0^-\ar@<1ex>[r]^-{\mathcal{R}_-}
      &\ar[l]^-{\mathcal{R}_+}\psi_1^-\ar@<1ex>[r]^-{\mathcal{R}_-}
      &\ar[l]^-{\mathcal{R}_+}\psi_2^-\ar@<1ex>[r]^-{\mathcal{R}_-}
      &\ar[l]^-{\mathcal{R}_+}\ \dots&
      \\
      \dots\ \ar@<1ex>[r]^-{\mathcal{R}_-}
      &\ar[l]^-{\mathcal{R}_+}\xi_{2}^+\ar@<1ex>[r]^-{\mathcal{R}_-}
      &\ar[l]^-{\mathcal{R}_+}\xi_{1}^+\ar@<1ex>[r]^-{\mathcal{R}_-}
      &\ar[l]^-{\mathcal{R}_+}0^\pm\ar@<1ex>[r]^-{\mathcal{R}_-}
      &\ar[l]^-{\mathcal{R}_+}\theta_{3}^\pm\ar@<1ex>[r]^-{\mathcal{R}_-}
      &\ar[l]^-{\mathcal{R}_+}\theta_{2}^\pm\ar@<1ex>[r]^-{\mathcal{R}_-}
      &\ar[l]^-{\mathcal{R}_+}\theta_{1}^\pm\ar@<1ex>[r]^-{\mathcal{R}_-}
      &\ar[l]^-{\mathcal{R}_+}\theta_{0}^\pm\ar@<1ex>[r]^-{\mathcal{R}_-}
      &\ar[l]^-{\mathcal{R}_+}\omega_{1}^-\ar@<1ex>[r]^-{\mathcal{R}_-}
      &\ar[l]^-{\mathcal{R}_+}\omega_{2}^-\ar@<1ex>[r]^-{\mathcal{R}_-}
      &\ar[l]^-{\mathcal{R}_+}\ \dots
    }
  \end{equation*}
  \caption{The Pavlov equation: action of recursion operators
    \eref{direct_recursion_operator_for_Pavlov_eq},
    \eref{inverse_recursion_operator_for_Pavlov_eq}}
  \label{tab:RO-act-Pavlov_eq}
\end{figure}

\section{The universal hierarchy equation}
\label{sec:univ-hier-equat}

The universal hierarchy equation (UHE) was discussed
in~\cite{MartinezAlonsoShabat2002,MartinezAlonsoShabat2004}.

\subsection{The equation}
\label{sec:equation-2}

The UHE  reads
\begin{equation}
  \label{eq:3}
  u_{yy}=u_t\,u_{xy}-u_y\,u_{tx}.
\end{equation}
We assign the weights $\abs{x}=0$, $\abs{y}=1$, $\abs{t}=0$, $\abs{u}=-1$ to
the variables~$x$, $y$, $t$, and~$u$.

Similar to \Sref{sec:equation-1}, we consider the internal coordinates
\begin{equation*}
  u_{k,l}^0=u_{\underbrace{x\dots x}_{k\textup{\tiny\ times}}\underbrace{t\dots
      t}_{l\textup{\tiny\ times}}},\quad u_{k,l}^1=u_{\underbrace{x\dots x}_{k\textup{\tiny\
        times}}
    \underbrace{t\dots
      t}_{l\textup{\tiny\ times}}},  \qquad k,l\geq 0.
\end{equation*}
on~$\mathcal{E}$. Consequently, $\abs{u_{k,l}^0} = -1$,
$\abs{u_{k,l}^1} = -2$.

The total derivatives in the chosen coordinates are
\begin{eqnarray*}
  D_x&=\frac{\partial}{\partial
       x}+\sum_{k,l}\left(u_{k+1,l}^0\frac{\partial}{\partial
       u_{k,l}^0} + u_{k+1,l}^1\frac{\partial}{\partial u_{k,l}^1}\right),\\
  D_y&=\frac{\partial}{\partial
       y}+\sum_{k,l}\left(u_{k,l}^1\frac{\partial}{\partial
       u_{k,l}^0} +
       D_x^kD_t^l\left(u_{01}^0u_{10}^1-
       u_{00}^1u_{11}^0\right)\frac{\partial}{\partial
       u_{k,l}^1}\right),\\
  D_t&=\frac{\partial}{\partial
       t}+\sum_{k,l}\left(u_{k,l+1}^0\frac{\partial}{\partial
       u_{k,l}^0} + u_{k,l+1}^1\frac{\partial}{\partial u_{k,l}^1}\right).
\end{eqnarray*}

Local symmetries of~$\mathcal{E}$ are solutions to the equation
$\ell_{\mathcal{E}}(\phi)\equiv
D_y^2(\phi)-u_tD_xD_y(\phi)+u_yD_xD_t(\phi)-u_{xy}D_t(\phi) +u_{xt}
D_y(\phi)=0$.
The space~$\sym(\mathcal{E})$ is spanned by the functions
$\theta_0(X)=Xu_x-X'u$, $\theta_1(X)=X$, $\phi_0(T)=Tu_t+T'yu_y$,
$\phi_1(T)=Tu_y$, $\upsilon=yu_y+u$, where~$X$ is a function of~$x$ and $T$ is
a function of $t$, while `prime' denotes the corresponding derivatives. The
commutators are presented in \Tref{tab:UHE-loc-symm-comm}.
\begin{table}
  \caption{\label{tab:UHE-loc-symm-comm}The UHE: commutators of local
    symmetries.}
\begin{indented}
\item[]\begin{tabular}{@{}l|c|c|c|c|c}
\br
    \rule{0em}{10pt}&$\upsilon$
    &$\theta_0(\bar{X})$
    &$\theta_1(\bar{X})$
    &$\phi_0(\bar{T})$
    &$\phi_1(\bar{T})$
    \\\hline
    \rule{0em}{10pt}$\upsilon$
    &$0$
    &$0$
    &$-\theta_1(\bar{X})$
    &$0$
    &$\phi_1(\bar{T})$
    \\\hline
    \rule{0em}{10pt}$\theta_0(X)$
    &\dots&$\theta_0([\bar{X},X])$
    &$\theta_1([\bar{X},X])$
    &$0$
    &$0$
    \\\hline
    \rule{0em}{10pt}$\theta_1(X)$&\dots&\dots&0&0&$0$
    \\\hline
    \rule{0em}{10pt}$\phi_0(T)$
    &\dots
    &\dots
    &\dots
    &$\phi_0([\bar{T},T])$
    &$\phi_1([\bar{T},T])$
    \\\hline
    \rule{0em}{10pt}$\phi_1(T)$
    &\dots
    &\dots
    &\dots
    &\dots
    &$0$\\
\br
\end{tabular}
\end{indented}
\end{table}
Weights of the evolutionary vector fields are
$\abs{\Ev_\upsilon}=\abs{\Ev_{\theta_0(X)}}=\abs{\Ev_{\phi_0(T)}}=0$,
$\abs{\Ev_{\theta_1(X)}} = 1$, $\abs{\Ev_{\phi_1(T)}}=-1$.

\subsection{The Lax pair and hierarchies}
\label{sec:lax-pair-hierarchies-2}

The UHE admits the following Lax representation
$q_{t}=\lambda^{-2}(\lambda u_t-u_y)q_x$, $q_{y}=\lambda^{-1}u_yq_{x}$.
Expansion in powers of~$\lambda$ leads to the system
$q_{i,t}=u_tq_{i+1,x}-u_yq_{i+2,x}$, $q_{i,y}=u_yq_{i+1,x}$.  The
corresponding positive covering is of the form
\begin{equation*}
  \begin{array}{lclclcl}
    q_{1,y}&=&\frac{u_t}{u_y},&&q_{1,x}&=&\frac{1}{u_y};\\[3mm]
    q_{i,y}&=&\frac{u_t}{u_y}q_{i-1,y}-q_{i-1,t},&&q_{i,x}&=&\frac{q_{i-1,y}}{u_y},
  \end{array}
\end{equation*}
$i>1$, with the additional variables~$q_i^{(j)}$ that satisfy the relations
$q_i^{(0)}=q_i$, $q_i^{(j+1)}=q_{i,t}^{(j)}$.  One has~$\abs{q_i^{(j)}}=i+1$.

The equations defining the negative covering are
\begin{equation*}
  \begin{array}{lclclcl}
    r_{1,y}&=&u_xu_y,&&r_{1,t}&=&u_xu_t-u_y;\\
    r_{i,y}&=&u_yr_{i-1,x},&&r_{i,t}&=&u_tr_{i-1,x}-r_{i-1,y},
  \end{array}
\end{equation*}
$i>1$, with $r_i^{(j)}$ defined by $r_i^{(j+1)}=r_{i,x}^{(j)}$.  The weights
are~$\abs{r_i^{(j)}}=-i-1$.

\subsection{Nonlocal symmetries in the positive covering}
\label{sec:nonl-symm-posit-2}

\subsubsection{Lifts of local symmetries}
\label{sec:lifts-local-symm-6}

The local symmetries of the UHE  are lifted as follows.

The symmetry~$\upsilon=yu_y+u$ is lifted to
$\Upsilon=(\upsilon,\upsilon^1,\upsilon^2,\dots,\upsilon^i,\dots)$,
where~$\upsilon^i= -(i + 1)q_i + yq_{i,y}$.  The lift
$\Theta_0(X)=(\theta_0,\theta_0^1,\theta_0^2,\dots,\theta_0^i,\dots)$
of~$\theta_0(X)=Xu_x - X' u$ is defined by
\begin{equation*}
\theta_0^i=\frac{X}{u_y}q_{i-1,y}.
\end{equation*}

Let us now introduce the operator
\begin{equation*}
  \mathcal{Y} = -y\frac{\partial}{\partial t} +
  2q_1\frac{\partial}{\partial y}  +
  \sum_{k=1}^\infty(k+2)q_{k+1}\frac{\partial}{\partial q_k}
\end{equation*}
and define by induction the quantities~$R_i(T)$ as follows:
\begin{equation}\label{eq:22}
  R_1(T)=-T'y,\qquad R_i(T)=\frac{1}{i}\mathcal{Y}(R_{i-1}(T)),\quad i\geq2.
\end{equation}
Then the lift of~$\phi_0(T)=Tu_t+T'yu_y$ is
$\Phi_0(T)=(\phi_0,\phi_0^1,\phi_0^2,\dots,\phi_0^i,\dots)$, where
$\phi_0^i=T q_{i,t} + T' yq_{i,y} + R_{i+1}(T)$, while the
symmetry~$\phi_1(T)=Tu_y$ is lifted by
$\Phi_1(T) = (\phi_1,\phi_1^1,\phi_1^2,\dots,\phi_1^i,\dots)$
with~$\phi_1^i=Tq_{i,y}-R_i(T)$.  Finally, the lift of~$\theta_1(X)=X$ is
$\Theta_1(X)=(\theta_1,0,\dots,0,\dots)$.

\subsubsection{Nonlocal symmetries}
\label{sec:nonlocal-symmetries-6}

There exists a family of invisible symmetries
\begin{equation*}
  \Phi_\inv^i(T) =(\underbrace{0,\dots,0}_{i \textup{\tiny\ times}},\phi_\inv^1,\dots,
  \phi_\inv^k,\dots),
\end{equation*}
where~$\phi_\inv^1=T$ and $\phi_\inv^k=R_{i-1}(T)$, $i>1$, $R_{i-1}(T)$ being
defined by \Eref{eq:22}.

The UHE also admits another two families of nonlocal symmetries in~$\tau^+$
defined as follows. Let us set
$\Psi_0=(\psi_0,\psi_0^1,\psi_0^2,\dots,\psi_0^i,\dots)$,
where~$\psi_0= 2q_1u_y - yu_t$ and
$\psi_0^i= -(i + 2)q_{i+1} - yq_{i,t} + 2q_1 q_{i,y}$.  We also introduce
$\Psi_1=(\psi_1,\psi_1^1,\psi_1^2,\dots,\psi_1^i,\dots)$ with
$\psi_1= -3q_2 u_y + 2q_1 u_t - yu_y q_{1,t}$ and
$\psi_1^i= (i + 3)q_{i+2} + yq_{i+1,t} + 2q_1 q_{i,t} - (3q_2 +
yq_{1,t})q_{i,y}$.
Then we set $\Psi_k=[\Psi_0,\Psi_{k-1}]$, $k\geq2$, and
$\Xi_l(T)=[\Psi_l,\Phi_1(T)]$, $l\geq1$.  Distribution of the constructed
symmetries along weights is given by
$\abs{\Upsilon}=\abs{\Theta_0(X)}=\abs{\Phi_0(T)}=0$, $\abs{\Theta_1(X)} = 1$,
$\abs{\Phi_1(T)}=-1$, $\abs{\Psi_k}=k+1$, $k\geq 0$,
$\abs{\Phi_l^\inv(T)}=-l-1$, $\abs{\Xi_l(T)} = l$, $l\geq 1$.

\subsubsection{Lie algebra structure}
\label{sec:lie-algebra-struct-6}

Consider the following subspaces in~$\sym_{\tau^+}(\mathcal{E})$: $W$ spanned
by $\Upsilon$, $\Psi_i$, $i\geq0$, $V[x]$ spanned by $\Theta_0(X)$,
$\Theta_1(X)$, $V[t]$ spanned by $\Phi_0(T)$, $\Phi_1(T)$, $\Phi_\inv^i(T)$,
and $\Xi_j(T)$, $i$, $j\geq1$.  Then we have the following
\begin{theorem}
  \label{thm:UHE+}
  There exist bases $\mathbf{w}_i$\textup{,} $i\geq0$\textup{,} in
  $W$\textup{,} $\mathbf{v}_0(X)$\textup{,} $\mathbf{v}_1(X)$\textup{,} in
  $V[x]$\textup{,} $\mathbf{v}_i(T)$\textup{,} $i\in\mathbb{Z}$\textup{,} in
  $V[t]$\textup{,} such that their commutators satisfy the relations presented
  in \Tref{tab:UHE:comm-in-tau+}.  Thus\textup{,} $\sym_{\tau^+}(\mathcal{E})$
  is isomorphic to
  $\mathfrak{W}_0^+\ltimes(\mathfrak{L}_2^+[x]\oplus\mathfrak{L}[t])$ with the
  natural action of ~$\mathfrak{W}_0^+$ on~$\mathfrak{L}_2^+[x]$
  and~$\mathfrak{L}[t]$.
\end{theorem}
\begin{table}
\caption{\label{tab:UHE:comm-in-tau+}The UHE: commutators in
  $\sym_{\tau^+}(\mathcal{E})$.}
\begin{indented}
\item[]\begin{tabular}{@{}l|c|c|c}
\br
    \rule{0em}{10pt}
    &$\mathbf{w}_j$&$\mathbf{v}_j(\bar{X})$&$\mathbf{v}_j(\bar{T})$\\\hline
    \rule{0em}{17pt}$\mathbf{w}_i$
    &$(j-i)\mathbf{w}_{i+j}$&$\begin{array}{ll}
                                j\mathbf{v}_{i+j}(\bar{X}),&0\leq i+j\leq 1,\\
                                0,&\textup{otherwise}\end{array}$
                   &$j\mathbf{v}_{i+j}(\bar{T})$\\\hline
    \rule{0em}{17pt}$\mathbf{v}_i(X)$
    &\dots&$\begin{array}{ll}\mathbf{v}_{i+j}([X,\bar{X}]),&0\leq i+j\leq 1,\\
              0,&\textup{otherwise}\end{array}$&$0$\\\hline
    \rule{0em}{10pt}$\mathbf{v}_i(T)$
    &\dots&\dots&$\mathbf{v}_{i+j}([T,\bar{T}])$\\
\br
\end{tabular}
\end{indented}
\end{table}

%%%%%%%%%%%%%%%%%%%%%%%%%%%%%%%

\subsection{Nonlocal symmetries in the negative covering}
\label{sec:nonl-symm-negat-2}

\subsubsection{Lifts of local symmetries}
\label{sec:lifts-local-symm-7}

The symmetry~$\upsilon = yu_y + u$ is lifted to
$\Upsilon=(\upsilon,\upsilon^1,\upsilon^2,\dots,\upsilon^i,\dots)$,
where~$\upsilon^i= (i + 1)r_i + yu_y r_{i-1,x}$ and $r_0$ denotes~$u$.  The
lift of~$\theta_0(X) = Xu_x - X'u$ is
$\Theta_0(X)=(\theta_0(X),\theta_0^1,\theta_0^2,\dots,\theta_0^i,\dots)$ with
$\theta_0^i= X r_{i,x} - R_{i+1}(X)$, $R_{i+1}$ being defined by \Eref{eq:22}.
For the sym\-met\-ry~$\phi_0(T)= T u_t + T' yu_y$ one has
$\Phi_0(T)=(\phi_0(T),\phi_0^1,\phi_0^2,\dots,\phi_0^i,\dots)$,
where~$\phi_0^i= T r_{i,t} + T' yu_y r_{i-1,x}$.  The
symmetry~$\phi_1(T) = Tu_y$ is lifted to
$\Phi_1(T)=(\phi_1(T),\phi_1^1,\phi_1^2,\dots,\phi_1^i,\dots)$,
where~$\phi_1^i=Tr_{i,y}$.  Finally, for~$\theta_1(X)=X$ one has
$\Theta_1(X)=(\theta_1(X),R_1(X),\dots,R_i(X),\dots)$, where again $R_i$ is
defined by~\eref{eq:22}.

\subsubsection{Nonlocal symmetries}
\label{sec:nonlocal-symmetries-7}

In the $\tau^-$ covering of the UHE  there exists a family of
invisible symmetries of the form
\begin{equation*}
  \Phi_\inv^k(X)= (\underbrace{0,\dots,0}_{k \textup{\tiny\
      times}},\phi_\inv^1,\dots,\phi_\inv^i,\dots),
\end{equation*}
where~$\phi_\inv^1=X$ and~$\phi_\inv^i=R_{i-1}(X)$ (see \Eref{eq:22} for the
definition of~$R_i$).

Consider now two nonlocal symmetries
$\Psi_j=(\psi_j,\psi_j^1,\psi_j^2,\dots,\psi_j^i,\dots)$, $j=-1$, $-2$, defined
by~$\psi_{-1}= 2r_1 - uu_x$, $\psi_{-1}^i= (i + 2)r_{i+1}- ur_{i,x}$, and
$\psi_{-2}= 3r_2 - 2r_1 u_x - ur_{1,x} + uu_x^2$,
$\psi_{-2}^i= (i + 3)r_{i+2} - ur_{i+1,x} + (uu_x - 2r_1 )r_{i,x}$.  We now
introduce two families of nonlocal symmetries by setting
$\Psi_{-k}=[\Psi_{-1},\Psi_{-k+1}]$, $k\leq-3$, $\Omega_l(X)=[\Psi_l,\Phi_5(X)]$,
$l\leq -1$.

The~$\tau^-$-nonlocal symmetries are distributed along weights as follows:
$\abs{\Upsilon}=\abs{\Theta_0(X)}=\abs{\Phi_0(T)}=0$, $\abs{\Phi_1(T)}=-1$,
$\abs{\Theta_1(X)}=1$, $\abs{\Psi_k}=-k$, $k\leq-1$,
$\abs{\Phi_i^\inv(X)}=i+1$, $i\geq 1$, $\abs{\Omega_j(X)} =j$, $j\leq-1$.

\subsubsection{Lie algebra structure}
\label{sec:lie-algebra-struct-7}

Consider the following subspaces in~$\sym_{\tau^-}(\mathcal{E})$: $W$ spanned
by $\Upsilon$, $\Psi_k$, $k\leq-1$, $V[x]$ spanned by $\Omega_l(X)$, $l\geq1$,
$\Theta_0(X)$, $\Theta_1(X)$, and $\Phi_\inv^k(X)$, $k\geq1$, $V[t]$ spanned
by $\Phi_0(T)$, $\Phi_1(T)$.  Then the following result holds:

\begin{theorem}
  \label{thm:UHE-}
  There exist bases $\mathbf{w}_i$\textup{,} $i\leq0$\textup{,} in
  $W$\textup{,} $\mathbf{v}_i(X)$\textup{,} $i\in\mathbb{Z}$\textup{,} in
  $V[x]$\textup{,} $\mathbf{v}_i(T)$\textup{,} $i=0$\textup{,} $-1$\textup{,}
  in $V[t]$\textup{,} such that their commutators satisfy the relations
  presented in \Tref{tab:UHE:comm-in-tau-}.  Thus\textup{,}
  $\sym_{\tau^-}(\mathcal{E})$ is isomorphic to
  $\mathfrak{W}_0^-\ltimes(\mathfrak{L}_2^-[t]\oplus\mathfrak{L}[x])$ with the
  natural action of ~$\mathfrak{W}_0^-$
  on~$\mathfrak{L}_2^-[t]\oplus\mathfrak{L}[x]$.
\end{theorem}
\begin{table}
\caption{\label{tab:UHE:comm-in-tau-}The UHE: commutators in
  $\sym_{\tau^-}(\mathcal{E})$.}
\begin{indented}
\item[]\begin{tabular}{@{}l|c|c|c}
\br
    \rule{0em}{10pt}
    &$\mathbf{w}_j$&$\mathbf{v}_j(\bar{X})$&$\mathbf{v}_j(\bar{T})$\\\hline
    \rule{0em}{17pt}$\mathbf{w}_i$
    &$(j-i)\mathbf{w}_{i+j}$
                   &$j\mathbf{v}_{i+j}(\bar{X})$
                                           &$\begin{array}{ll}
                                               j\mathbf{v}_{i+j}(\bar{T}),
                                               &-1\leq i+j\leq0,\\
                                               0,&\textup{otherwise}
                                             \end{array}
                                                                                               $\\\hline
    \rule{0em}{11pt}$\mathbf{v}_i(X)$
    &\dots&$\mathbf{v}_{i+j}([X,\bar{X}])$&$0$\\\hline
    \rule{0em}{10pt}$\mathbf{v}_i(T)$
    &\dots&\dots&$\begin{array}{ll}
                    \mathbf{v}_{i+j}([T,\bar{T}])&-1\leq i+j\leq0,\\
                    0,&\textup{otherwise}
                  \end{array}
                        $\\
\br
\end{tabular}
\end{indented}
\end{table}

\subsection{Recursion operators}
\label{sec:recursion-operators-2}

The following proposition describes recursion operators for the symmetries of
the~UHE (see~\cite{Morozov2014}):

\begin{proposition}
  \Eref{eq:3} admits the recursion operator for symmetries
  $\psi=\mathcal{R}_{+}(\varphi)$ defined by the following system\textup{:}
  \begin{equation}
    \begin{array}{l}
    D_x(\psi) = u_y^{-1}\,\left(-D_y(\varphi)+u_{xy}\,\psi\right),
    \\
    D_y(\psi) = D_t(\varphi)-u_y^{-1}\,\left(u_t\,D_y(\varphi)+(u_y\,u_{tx}-
      u_t\,u_{xy})\,\psi\right).
    \end{array}
    \label{direct_recursion_operator_for_UHE}
  \end{equation}
  The inverse operator $\varphi=\mathcal{R}_{-}(\psi)$ is defined by system
  \begin{equation}\fl
    D_t(\varphi) = D_y(\psi)-u_t\,D_x(\psi)+u_{tx}\,\psi,
    \quad
    D_y(\varphi) = -u_y\,D_x(\psi)+u_{xy}\,\psi.
    \label{inverse_recursion_operator_for_UHE}
  \end{equation}
\end{proposition}
The action of the recursion operators on local symmetries and shadows is
schematically shown in \Fref{tab:RO-act-UHE}.
\begin{figure}[b]
  \begin{equation*}\xymatrixcolsep{1.3pc}\xymatrixrowsep{1.3pc}\fl
    \xymatrix{
      &\dots\ \ar@<1ex>[r]^-{\mathcal{R}_-}
      &\ar[l]^-{\mathcal{R}_+}\psi^{+}_{2}\ar@<1ex>[r]^-{\mathcal{R}_-}
      &\ar[l]^-{\mathcal{R}_+}\psi^{+}_{1}\ar@<1ex>[r]^-{\mathcal{R}_-}
      &\ar[l]^-{\mathcal{R}_+}\psi^{+}_{0}\ar@<1ex>[r]^-{\mathcal{R}_-}
      &\ar[l]^-{\mathcal{R}_+}\upsilon^\pm\ar@<1ex>[r]^-{\mathcal{R}_-}
      &\ar[l]^-{\mathcal{R}_+}\psi^{-}_{-1}\ar@<1ex>[r]^-{\mathcal{R}_-}
      &\ar[l]^-{\mathcal{R}_+}\psi^{-}_{-2}\ar@<1ex>[r]^-{\mathcal{R}_-}
      &\ar[l]^-{\mathcal{R}_+}\psi^{-}_{-3}\ar@<1ex>[r]^-{\mathcal{R}_-}
      &\ar[l]^-{\mathcal{R}_+}\ \dots&
      \\
      \dots\ \ar@<1ex>[r]^-{\mathcal{R}_-}
      &\ar[l]^-{\mathcal{R}_+}\xi^{+}_{2}\ar@<1ex>[r]^-{\mathcal{R}_-}
      &\ar[l]^-{\mathcal{R}_+}\xi^{+}_{1}\ar@<1ex>[r]^-{\mathcal{R}_-}
      &\ar[l]^-{\mathcal{R}_+}\phi_{0}^\pm\ar@<1ex>[r]^-{\mathcal{R}_-}
      &\ar[l]^-{\mathcal{R}_+}\phi_1^\pm\ar@<1ex>[r]^-{\mathcal{R}_-}
      &\ar[l]^-{\mathcal{R}_+}0^\pm\ar@<1ex>[r]^-{\mathcal{R}_-}
      &\ar[l]^-{\mathcal{R}_+}\theta_{1}^\pm\ar@<1ex>[r]^-{\mathcal{R}_-}
      &\ar[l]^-{\mathcal{R}_+}\theta_{0}^\pm\ar@<1ex>[r]^-{\mathcal{R}_-}
      &\ar[l]^-{\mathcal{R}_+}\omega^{-}_{1}\ar@<1ex>[r]^-{\mathcal{R}_-}
      &\ar[l]^-{\mathcal{R}_+}\omega^{-}_{2}\ar@<1ex>[r]^-{\mathcal{R}_-}
      &\ar[l]^-{\mathcal{R}_+}\ \dots
    }
  \end{equation*}
  \caption{The UHE: action of recursion operators
    \eref{direct_recursion_operator_for_UHE},
    \eref{inverse_recursion_operator_for_UHE}}
  \label{tab:RO-act-UHE}
\end{figure}

\section{The modified Veronese web equation}
\label{sec:modif-veron-web}

The modified Veronese web equation (mVWE) was studied in~\cite{A-Sh} and is
related to the Veronese web equation, \cite{V-Web-2,V-Web-1}, by the
B\"acklund transformation (\ref{eq:14}).

\subsection{The equation}
\label{sec:equation-3}

The mVWE has the form
\begin{equation}
  \label{eq:4}
  u_{ty}=u_tu_{xy}-u_yu_{tx}.
\end{equation}
We assign zero weights to all the variables under consideration. Internal
coordinates are chosen similar to the previous cases, i.e.,
\begin{equation*}
  u_k=u_{\underbrace{x\dots x}_{k \textup{\tiny\ times}}},\quad
  u_{k,l}^t=u_{\underbrace{x\dots x}_{k \textup{\tiny\ times}}\underbrace{t\dots t}_{l
      \textup{\tiny\ times}}},\quad
  u_{k,l}^y=u_{\underbrace{x\dots x}_{k \textup{\tiny\ times}}\underbrace{y\dots y}_{l
      \textup{\tiny\ times}}},
\end{equation*}
where $k\geq0$, $l>0$. Then the total derivatives read
\begin{eqnarray*}\fl
  D_x=\frac{\partial}{\partial x} + \sum_{k}u_{k+1}\frac{\partial}{\partial
       u_k} + \sum_{k,l}\left(
       u_{k+1,l}^y\frac{\partial}{\partial u_{k,l}^y} +
       u_{k+1,l}^t\frac{\partial}{\partial u_{k,l}^t}\right),\\\fl
  D_y=\frac{\partial}{\partial y} + \sum_ku_{k,1}^y\frac{\partial}{\partial
       u_k} +
       \sum_{k,l}\left(u_{k,l+1}^y\frac{\partial}{\partial u_{k,l}^y} +
       D_x^kD_t^{l-1}(u_{01}^tu_{11}^y - u_{01}^yu_{11}^t)\frac{\partial}{\partial
       u_{k,l}^t}\right),\\\fl
  D_t=\frac{\partial}{\partial t} + \sum_ru_{k,1}^t\frac{\partial}{\partial
       u_k} + \sum_{k,l}\left(D_x^kD_y^{l-1}(u_{01}^tu_{11}^y -
       u_{01}^yu_{11}^t)\frac{\partial}{\partial u_{k,l}^y} +
       u_{k,l+1}^t\frac{\partial}{\partial u_{k,l}^t}\right).
\end{eqnarray*}
Symmetries are defined by the equation
\begin{equation}\fl
  \label{eq:8}
  \ell_{\mathcal{E}}(\phi)\equiv D_tD_y(\phi)-u_tD_xD_y(\phi)+u_yD_tD_x(\phi)-
  u_{xy}D_t(\phi) + u_{tx}D_y(\phi)=0.
\end{equation}
The space of solutions is generated by the functions $\phi(T)=Tu_t$,
$\upsilon(Y)=Yu_y$, $\theta_0(X)=Xu_x-X'u$, $\theta_1(X)=X$, where~$X=X(x)$,
$Y=Y(y)$, and~$T=T(t)$ are arbitrary functions of their arguments.  The
commutators of the symmetries are presented in \Tref{tab:mVWE-loc-symm-comm}.
\begin{table}
\caption{\label{tab:mVWE-loc-symm-comm}The mVwe: commutators of local symmetries.}
\begin{indented}
\item[]\begin{tabular}{@{}l|c|c|c|c}
\br
    \rule{0em}{10pt}
    &$\phi(\bar{T})$
    &$\theta_0(\bar{X})$&$\theta_1(\bar{X})$&$\upsilon(\bar{Y})$
        \\\hline
    \rule{0em}{10pt}$\phi(T)$&$\phi([\bar{T},T])$&$0$&$0$&$0$
    \\\hline
    \rule{0em}{10pt}$\theta_0(X)$&\dots&$\theta_0([\bar{X},X])$&
             $\theta_1([\bar{X},X])$&$0$
    \\\hline
    \rule{0em}{10pt}$\theta_1(X)$&\dots&\dots&0&$0$
    \\\hline
    \rule{0em}{10pt}$\upsilon(Y)$&\dots&\dots&$\dots$&$\upsilon([\bar{Y},Y])$\\
\br
\end{tabular}
\end{indented}
\end{table}

\subsection{The Lax pair and hierarchies}
\label{sec:lax-pair-hierarchies-3}

The mVwe admits the Lax pair
\begin{equation}
  \label{eq:14}
  q_{t} = (\lambda+1)^{-1}u_tq_x,\qquad
  q_{y} = \lambda^{-1}u_yq_{x}.
\end{equation}
Expanding in powers of~$\lambda$, one obtains $q_{i-1,t}+q_{i,t}=u_tq_{i,x}$,
$q_{i-1,y}=u_yq_{i,x}$.  Then the positive covering acquires the form
\begin{eqnarray*}
  q_{1,t}&=\frac{u_t}{u_y},&\qquad q_{1,x}=\frac{1}{u_y},\\
  q_{i,x}&=\frac{q_{i-1,y}}{u_y},&\qquad q_{i,t}=\frac{u_t}{u_y}q_{i-1,y}-q_{i-1,t},
\end{eqnarray*}
$i>1$, the additional variables being~$q_i^{(j)}$ defined as usual:
$ q_i^{(0)}=q_i$, $q_i^{(j+1)}=q_{i,y}^{(j)}$ with~$\abs{q_i^{(j)}}=0$.

The defining equations for the negative covering are
\begin{eqnarray*}
  r_{1,t}&=u_t(u_x-1),&\qquad r_{1,y}=u_xu_y;\\
  r_{i,t}&=u_tr_{i-1,x}-r_{i-1,t},&\qquad r_{i,y}=u_yr_{i-1,x},
\end{eqnarray*}
$i>1$. The auxiliary variables are~$r_i^{(j)}$, defined by $r_i^{(0)}=r_i$,
$r_i^{(j+1)}=r_{i,y}^{(j)}$.  Similar to the positive case, their weights are
trivial.

\subsection{Nonlocal symmetries in the positive covering}
\label{sec:nonl-symm-posit-3}

\subsubsection{Lifts of local symmetries}
\label{sec:lifts-local-symm}

All the local symmetries can be lifted to the~$\tau^+$ covering. Namely, the
lift of~$\phi_1(T)=Tu_t$ is
$\Phi(T)=\left(\phi(T),\phi^1,\dots,\phi^i,\dots\right)$,
where~$\phi^i=Tq_{i,t}$.  The lift of~$\theta_0(X)=Xu_x-X'u$ is given by
$\Theta_0(X)=\left(\theta_0(X),\theta_0^1,\dots,\theta_0^i,\dots\right)$,
where~$\theta_0^i=Xq_{i,x}$. To lift the symmetry~$\upsilon(Y)=Yu_y$, consider
the operator
\begin{equation*}
  \mathcal{Y}=q_1\frac{\partial}{\partial y} + \sum_{k=1}^\infty
  q_{k+1}\frac{\partial}{\partial q_k}
\end{equation*}
and set recursively
\begin{equation}
  \label{eq:15}
  R_1(Y)=Y'q_1,\qquad R_n(Y)=\frac{1}{n}\mathcal{Y}(R_{n-1}).
\end{equation}
Then
$ \Upsilon(Y)=\left(\upsilon(Y),\upsilon^1,\dots,\upsilon^i,\dots\right)$,
where~$\upsilon^i=Yq_{i,y}-R_i(Y)$.  Finally, for $\theta_1(X)=X$ one has
$\Theta_1=(\theta_1(X),0,\dots,0,\dots)$ for the lift of~$\theta_1(X)=X$.

\subsubsection{Nonlocal symmetries}
\label{sec:nonlocal-symmetries}

There exist three families of `purely nonlocal' symmetries in~$\tau^+$. The
first consists of the invisible symmetries which are of the form
\begin{equation*}
  \Phi_\inv^k(Y)=\Big(\underbrace{0,\dots,0}_{k\textup{\tiny\
      times}},\phi_\inv^1,\dots, \phi_\inv^i,\dots\Big),
\end{equation*}
where~$\phi_\inv^1=Y$ and~$\phi_\inv^i=R_{i-1}(Y)$, $i>1$, $R_l(Y)$ being
defined by~\eref{eq:15}.

The second family is constructed as follows: symmetries~$\Psi_0$ and~$\Psi_1$
are defined by $\Psi_0=(\psi_0^0,\psi_0^1,\dots,\psi_0^i,\dots)$,
where~$\psi_0^0=q_1u_y+u$, $\psi_0^i=-(i+1)q_{i+1}-iq_i+q_1q_{1,y}$, $i>0$,
and $\Psi_1=(\psi_1^0,\psi_1^1,\dots,\psi_1^i,\dots)$,
where~$\psi_1^0=(-2q_2-q_1+q_1q_{1,y})u_y$,
$\psi_1^i=(i+2)q_{i+2}+(i+1)q_{i+1} -q_1q_{i+1,y} +
(-2q_2-q_1+q_1q_{1,y})q_{i,y}$,
$i>0$. Let us also set by induction
$\Psi_k=[\Psi_0,\Psi_{k-1}] + k\Psi_{k-1}$, $k>1$.

The third family consists of the symmetries
$\Xi_k(Y)=[\Psi_k,\Phi_\inv^1(Y)]-(k-1)!\Upsilon(Y)$, $k=0,1,\dots$

\subsubsection{Lie algebra structure}
\label{sec:lie-algebra-struct}

Consider the following subspaces in~$\sym_{\tau^+}(\mathcal{E})$: $V[x]$
spanned by $\Theta_0(X)$, $\Theta_1(X)$, $V[t]$ spanned by $\Phi(T)$, $V[y]$
spanned by $\Xi_k(Y)$, $\Upsilon(Y)$, and $\Phi_\inv^l(Y)$, $W$ spanned by
$\Psi_k$.  The following result holds:
\begin{theorem}
  \label{thm:mVWE+}
  There exist bases $\mathbf{w}_i$\textup{,} $i\geq 1$\textup{,} in
  $W$\textup{,} $\mathbf{v}_i(X)$\textup{,} $i=0$\textup{,} $1$\textup{,} in
  $V[x]$\textup{,} $\mathbf{v}_i(Y)$\textup{,} $i\in\mathbb{Z}$\textup{,} in
  $V[y]$\textup{,} $\mathbf{v}(T)$ in $V[t]$\textup{,} such that their
  commutators satisfy the relations presented in
  \Tref{tab:mVWE-nonloc-symm-comm+}.  In other words\textup{,}
  $\sym_{\tau^+}(\mathcal{E})$ is isomorphic to
  $\widetilde{\mathfrak{W}}_0^+
  \ltimes(\mathfrak{L}[y]\oplus\mathfrak{L}_2^+[x]) \oplus \mathfrak{V}[t]$
  with the natural action of the Witt algebra $\mathfrak{W}_0^+$
  on~$\mathfrak{L}[y]\oplus\mathfrak{L}_2^+[x]$.
  Here~$\widetilde{\mathfrak{W}}_0^+$ denotes the subalgebra
  in~$\mathfrak{W}_0^+$ generated by the
  elements~$\mathbf{e}_i-\mathbf{e}_0$\textup{,} $i\geq 1$.
\end{theorem}
\begin{table}
  \caption{\label{tab:mVWE-nonloc-symm-comm+}The mVwe: commutators in
    $\sym_{\tau^+}(\mathcal{E})$.}
  \begin{tabular}{@{}l|c|c|c|c|c}
    \br
    \rule{0em}{10pt}
    &$\mathbf{w}_j$
    &$\mathbf{v}_0(\bar{X})$
    &$\mathbf{v}_1(\bar{X})$
    &$\mathbf{v}_j(\bar{Y})$&$\mathbf{v}(\bar{T})$\\\hline
    \rule{0em}{10pt}$\mathbf{w}_i$
    &$\begin{array}{l}
        (j-i)\mathbf{w}_{i+j} +\\\qquad i\mathbf{w}_i-j\mathbf{w}_j
      \end{array}
    $&$0$&$-\mathbf{v}_1(\bar{X})$
    &$j(\mathbf{v}_{i+j}(\bar{Y})-\mathbf{v}_j(\bar{Y}))$&$0$\\\hline
    \rule{0em}{10pt}$\mathbf{v}_0(X)$
    &\dots&$\mathbf{v}_0([X,\bar{X}])$
    &$\mathbf{v}_1([X,\bar{X}])$&$0$&$0$\\\hline
    \rule{0em}{10pt}$\mathbf{v}_1(X)$&\dots&\dots&$0$&$0$&$0$\\\hline
    \rule{0em}{10pt}$\mathbf{v}_i(Y)$
    &\dots&\dots&\dots&$\mathbf{v}_{i+j}([Y,\bar{Y}])$&\\\hline
    \rule{0em}{10pt}$\mathbf{v}(T)$
    &\dots&\dots&\dots&\dots&$\mathbf{v}([T,\bar{T}])$\\
    \br
  \end{tabular}
\end{table}
\begin{remark}
  Lie algebra $\widetilde{\mathfrak{W}}_0^+$ is an example of a two-point
  Krichever-Novikov type algebra\textup{,}
  \textup{\cite{Schlichenmaier2014}}. In the base
  $\tilde{\mathbf{e}}_i =\mathbf{e}_i-\mathbf{e}_{i-1}=
  z^i\,(z-1)\,\partial/{\partial z}$\textup{,}
  $i \ge 1$\textup{,} it has an almost-graded structure
  $[\tilde{\mathbf{e}}_i,\tilde{\mathbf{e}}_j] =
  (j-i)\,(\tilde{\mathbf{e}}_{i+j} - \tilde{\mathbf{e}}_{i+j-1})$.
\end{remark}

\subsection{Nonlocal symmetries in the negative covering}
\label{sec:nonl-symm-negat-3}

\subsubsection{Lifts of local symmetries}
\label{sec:lifts-local-symm-1}

The symmetry $\phi(T)=Tu_t$ is lifted to
$\Phi(T)=(\phi(T),\phi^1,\dots,\phi^i,\dots)$, where~$\phi^i=Tr_{i,t}$.  To
define the lift of $\theta_0(X)=Xu_x-X'u$, consider the operator
\begin{equation*}
  \mathcal{Y}=u\frac{\partial}{\partial x} + 2r_1\frac{\partial}{\partial u} +
  \sum_{k=1}^\infty (k+2)r_{k+1}\frac{\partial}{\partial r_k}
\end{equation*}
and define by induction the quantities~$R_n(X)$ by setting
\begin{equation}
  \label{eq:16}
  R_1(X)=X'u,\qquad R_n(X)=\frac{1}{n}\mathcal{Y}(R_{n-1}).
\end{equation}
Then $\Theta_0(X)=(\theta_0(X),\theta_0^1,\dots,\theta_0^i,\dots)$,
where~$\theta_0^i=Xr_{i,x}-R_{i+1}(X)$, $R_n$ being defined by~\eref{eq:16}.
For $\upsilon(Y)=Yu_y$, one has
$ \Upsilon(Y)=(\upsilon,\upsilon^1,\dots,\upsilon^i,\dots)$
with~$ \upsilon^i=Yr_{i,y}$.  The symmetry $\theta_1(X)=X$ is lifted to
$\Theta_1(X)=(\theta_1,\theta_1^1,\dots,\theta_1^i,\dots)$,
where~$\theta_1^i=R_i(X)$.

\subsubsection{Nonlocal symmetries}
\label{sec:nonlocal-symmetries-1}

Similar to the positive case, three families of nonlocal symmetries arise
in~$\tau^-(\mathcal{E})$. The first consists of invisible symmetries
\begin{equation*}
  \Phi_\inv^k(X)=\Big(\underbrace{0,\dots,0}_{k\textup{\tiny\
      times}},\phi_\inv^1,\dots, \phi_\inv^i,\dots\Big),
\end{equation*}
where~$\phi_\inv^1=X$ and~$\phi_\inv^i=R_{i-1}(X)$, $i \geq2$.

Two nonlocal symmetries,
$\Psi_{-1}=(\psi_{-1},\psi_{-1}^1,\dots,\psi_{-1}^i,\dots)$ and
$\Psi_{-2}=(\psi_{-2},\psi_{-2}^1,\dots,\psi_{-2}^i,\dots)$, are constructed
explicitly. Namely, we set~$\psi_{-1}=2r_1-uu_x+u$,
$\psi_{-1}^i=(i+2)r_{i+1}+(i+1)r_i-ur_{i,x}$
and~$ \psi_{-2}=3r_2-2r_1u_x-ur_{1,x}+ uu_x^2-u$,
$\psi_{-2}^i = (i+3)r_{i+2} - (i+1)r_i -ur_{i+1,x} +(uu_x-2r_1)r_{i,x}$.  Then
the second family is defined by
$\Psi_{-k-1}=[\Psi_{-1},\Psi_{-k}]-k\Psi_{-k}+(-1)^{k+1}(k-3)!\,\Psi_{-1}$, $k>1$,
while the third one is
$\Omega_{-l}(X)=[\Psi_{-l},\Phi_4(X)]+(-1)^{l+1}(l-2)!\Theta_1(X)$, $l\geq 0$.

\subsubsection{Lie algebra structure}
\label{sec:lie-algebra-struct-1}

Let $W$ spanned by $\Psi_k$, $V[x]$ spanned by $\Omega_l(X)$, $\Theta_0(X)$,
$\Theta_1(X)$, and $\Phi_\inv^i(X)$, $V[t]$ spanned by $\Phi(T)$, $V[y]$
spanned by $\Upsilon(Y)$ be subspaces in~$\sym_{\tau^-}(\mathcal{E})$.
\begin{theorem}
  \label{thm:mVWE-}
  There exist bases $\mathbf{w}_i$\textup{,} $i\leq-1$\textup{,} in
  $W$\textup{,} $\mathbf{v}_i(X)$\textup{,} $i\in\mathbb{Z}$\textup{,} in
  $V[x]$\textup{,} $\mathbf{v}(T)$ in $V[t]$\textup{,} $\mathbf{v}(Y)$ in
  $V[y]$\textup{,} such that their commutators satisfy the relations presented
  in \Tref{tab:mVWE-nonloc-symm-comm-}.  Thus\textup{,}
  $\sym_{\tau^-}(\mathcal{E})$ is isomorphic to
  $\widetilde{\mathfrak{W}}_0^- \ltimes\mathfrak{L}[x]\oplus \mathfrak{V}[y]
  \oplus \mathfrak{V}[t]$
  with the natural action of the Witt algebra $\mathfrak{W}$
  on~$\mathfrak{L}[x]$. Here~$\widetilde{\mathfrak{W}}_0^-$ denotes the
  subalgebra in~$\mathfrak{W}_0^-$ generated by the
  elements~$\mathbf{e}_i-\mathbf{e}_0$\textup{,} $i\leq -1$.
\end{theorem}
\begin{table}
\caption{\label{tab:mVWE-nonloc-symm-comm-}The mVwe: commutators in
  $\sym_{\tau^-}(\mathcal{E})$.}
\begin{indented}
\item[]\begin{tabular}{@{}l|c|c|c|c}
\br
    &$\mathbf{w}_j$
    &$\mathbf{v}_j(\bar{X})$
    &$\mathbf{v}(\bar{Y})$&$\mathbf{v}(\bar{T})$\\\hline
    \rule{0em}{10pt}$\mathbf{w}_i$
    &$(j-i)\mathbf{w}_{i+j}+i\mathbf{w}_i-j\mathbf{w}_j$
    &$j(\mathbf{v}_{i+j}(\bar{X})-\mathbf{v}_j(\bar{X}))$&$0$&$0$\\\hline
    \rule{0em}{10pt}$\mathbf{v}_i(X)$
    &\dots&$\mathbf{v}_{i+j}([X,\bar{X}])$&$0$&$0$\\\hline
    \rule{0em}{10pt}$\mathbf{v}(Y)$
    &\dots&\dots&$\mathbf{v}([Y,\bar{Y}])$&$0$\\\hline
    \rule{0em}{10pt}$\mathbf{v}(T)$
    &\dots&\dots&\dots&$\mathbf{v}([T,\bar{T}])$\\
\br
\end{tabular}
\end{indented}
\end{table}

\begin{remark}
  Obviously, the Lie algebra $\widetilde{\mathfrak{W}}_0^-$ is isomorphic to
  $\widetilde{\mathfrak{W}}_0^+$\textup{,} the isomorphism
  $\mathbf{e}_{-k}-\mathbf{e}_0 \mapsto -(\mathbf{e}_k-\mathbf{e}_0)$\textup{,}
  $k\ge 1$\textup{,} is given by the change of variable $z \mapsto z^{-1}$.
\end{remark}

\subsection{Recursion operators}
\label{sec:recursion-operators-3}

To construct a recursion operator for \Eref{eq:4} we use the techniques of
\cite{Sergyeyev2015},
cf.~\cite{MalykhNutkuSheftel2004,MarvanSergyeyev2012,MorozovSergyeyev2014,KruglikovMorozov2015}
also.  We find a shadow for \Eref{eq:4} in the covering~\eref{eq:14}. It is of
the form $s=H(q)\,q_x^{-1}$, where $H$ is an arbitrary function in $q$.  Since
System~\eref{eq:14} is invariant with respect to the transformation
$q \mapsto H(q)$, we put, without loss of generality,
$s=q_x^{-1}$. Differentiation of~\eref{eq:14} by~$x$ and substitution for
$q_x =s^{-1}$ gives another covering
\begin{equation}
  s_t =(\lambda+1)^{-1}\,(u_t\,s_x-u_{tx}\,s),
  \qquad
  s_y =\lambda^{-1}\,(u_y\,s_x-u_{xy}\,s)
  \label{mVwe_s_covering}
\end{equation}
for \Eref{eq:4}.  Note that $s$ is a solution to the linearization~\eref{eq:8}
of \Eref{eq:4}.  Now put
\begin{equation}
  s = \sum \limits_{n=-\infty}^{\infty} s_n\, \lambda^n.
  \label{s_expansion}
\end{equation}
Since~\eref{eq:8} is independent of $\lambda$, each $s_n$ is a solution
to~\eref{eq:8}.  Substituting~\eref{s_expansion} to~\eref{mVwe_s_covering}
yields $s_{n-1,t} +s_{n,t}=u_ts_{n,x}-u_{tx}s_n$,
$s_{n-1,y} =u_y,s_{n,x}-u_{xy}s_n$.  Denoting $s_{n-1}=\phi$ and
$s_n =\psi$, we have
\begin{proposition}
  \label{thm:recursion_operator_for_mVwe}
  The system
  \begin{equation}
    \begin{array}{l}
      D_t(\psi) =
      -D_t(\varphi)+u_y^{-1}\,\left(u_t\,D_y(\varphi)+
      (u_t\,u_{xy}-u_y\,u_{tx})\,\psi\right),\\
      D_x(\psi) =u_y^{-1}\,\left(D_y(\varphi)+u_y\,\psi\right)
    \end{array}
    \label{direct_recursion_operator_for_mVwe}
  \end{equation}
  defines a recursion operator $\psi=\mathcal{R}_{+}(\varphi)$ for symmetries
  of \Eref{eq:4}.  The inverse operator $\varphi=\mathcal{R}_{-}(\psi)$ is
  given by system
  \begin{equation}\fl
    D_t(\varphi) = -D_t(\psi)+u_t\,D_x(\psi)-u_{tx}\,\psi,
    \quad
    D_y(\varphi) =u_y\,D_x(\psi)-u_{xy}\,\psi.
    \label{inverse_recursion_operator_for_mVwe}
  \end{equation}
\end{proposition}
The action of recursion operators $\mathcal{R}_{+}$ and $\mathcal{R}_{-}$ on
the shadows of nonlocal symmetries is more complicated than in the previous
sections. It is described by
\begin{proposition}
  The action of \eref{direct_recursion_operator_for_mVwe} and
  \eref{inverse_recursion_operator_for_mVwe} on the shadows
  $\psi_i^{\pm}$\textup{,} $\xi_i^{+}$\textup{,} $\omega_i^{-}$ is of the form
  \begin{equation}
    \mathcal{R}_{+}(\psi_i^{+}) = \sum \limits_{j=1}^{i+1} \alpha_{ij} \,\psi_j^{+},
    \quad \alpha_{i\,i+1} \neq 0,
    \quad i \ge 0,
    \label{R_plus_on_psi}
  \end{equation}
  \begin{equation}
    \mathcal{R}_{+}(\xi_i^{+}) = \sum \limits_{j=1}^{i+1} \beta_{ij} \,\xi_j^{+},
    \quad \beta_{i\,i+1} \neq 0,
    \quad i \ge 0,
    \label{R_plus_on_xi}
  \end{equation}
  \begin{equation}
    \mathcal{R}_{-}(\psi_{-k}^{-}) = \sum \limits_{j=1}^{k+1} \gamma_{kj}
    \,\psi_{-j}^{-}, 
    \quad \gamma_{k\,k+1} \neq 0,
    \quad k \ge 1,
    \label{R_minus_on_psi}
  \end{equation}
  \begin{equation}
    \mathcal{R}_{-}(\omega_i^{-}) = \sum \limits_{j=0}^{i+1} \delta_{ij}
    \,\omega_j^{-} 
    +\varepsilon_i \,\theta_0^{-},
    \quad \delta_{i\,i+1} \neq 0,
    \quad i \ge 0,
    \label{R_minus_on_omega}
  \end{equation}
  where $\alpha_{ij}$\textup{,} $\beta_{ij}$\textup{,} $\gamma_{kj}$\textup{,}
  and $\varepsilon_i$ are certain constants.  To find the action of
  $\mathcal{R}_{-}$ to $\psi_i^{+}$\textup{,} $\xi_i^{+}$ one has to apply
  $\mathcal{R}_{-}$ to both sides of \eref{R_plus_on_psi}\textup{,}
  \eref{R_plus_on_xi} and then to solve the obtained triangular systems. In
  the same way it is possible to find the action of $\mathcal{R}_{+}$ to
  $\psi_{-i}^{-}$, $\omega_i^{-}$.
\end{proposition}

These results are schematically shown in \Fref{tab:RO-act-mVwe}, where the
wavy arrows indicate the action
\eref{R_plus_on_psi}--\eref{R_minus_on_omega}. The usage of straight arrows
corresponds to Remark~\ref{sec:rddym-equat-synops-rem}.

\begin{figure}
  \begin{equation*}\xymatrixcolsep{1.5pc}\xymatrixrowsep{1.3pc}\fl
    \xymatrix{
      &\dots\ \ar@{~>}@<1ex>[r]^-{\mathcal{R}_-}
      &\ar@{~>}[l]^-{\mathcal{R}_+}\psi^{+}_{2}\ar@{~>}@<1ex>[r]^-{\mathcal{R}_-}
      &\ar@{~>}[l]^-{\mathcal{R}_+}\psi^{+}_{1}\ar@{~>}@<1ex>[r]^-{\mathcal{R}_-}
      &\ar@{~>}[l]^-{\mathcal{R}_+}\psi^{+}_{0}\ar@{->}@<1ex>[r]^-{\mathcal{R}_-}
      &\ar@{->}[l]^-{\mathcal{R}_+}\psi^{-}_{-1}\ar@{~>}@<1ex>[r]^-{\mathcal{R}_-}
      &\ar@{~>}[l]^-{\mathcal{R}_+}\psi^{-}_{-2}\ar@{~>}@<1ex>[r]^-{\mathcal{R}_-}
      &\ar@{~>}[l]^-{\mathcal{R}_+}\psi^{-}_{-3}\ar@{~>}@<1ex>[r]^-{\mathcal{R}_-}
      &\ar@{~>}[l]^-{\mathcal{R}_+}\ \dots&
      \\
      \dots\ \ar@{~>}@<1ex>[r]^-{\mathcal{R}_-}
      &\ar@{~>}[l]^-{\mathcal{R}_+}\xi^{+}_{1}\ar@{~>}@<1ex>[r]^-{\mathcal{R}_-}
      &\ar@{~>}[l]^-{\mathcal{R}_+}\xi^{+}_{0}\ar@<1ex>[r]^-{\mathcal{R}_-}
      &\ar[l]^-{\mathcal{R}_+}\upsilon^\pm\ar@<1ex>[r]^-{\mathcal{R}_-}
      &\ar[l]^-{\mathcal{R}_+}0^\pm\ar@<1ex>[r]^-{\mathcal{R}_-}
      &\ar[l]^-{\mathcal{R}_+}\theta_1^\pm\ar@<1ex>[r]^-{\mathcal{R}_-}
      &\ar[l]^-{\mathcal{R}_+}\theta_0^\pm\ar@<1ex>[r]^-{\mathcal{R}_-}
      &\ar[l]^-{\mathcal{R}_+}\omega^{-}_{0}\ar@{~>}@<1ex>[r]^-{\mathcal{R}_-}
      &\ar@{~>}[l]^-{\mathcal{R}_+}\omega^{-}_{1}\ar@{~>}@<1ex>[r]^-{\mathcal{R}_-}
      &\ar@{~>}[l]^-{\mathcal{R}_+}\ \dots
    }
  \end{equation*}
  \caption{The mVWE: action of recursion
    operators~\eref{direct_recursion_operator_for_mVwe},
    \eref{inverse_recursion_operator_for_mVwe}}
  \label{tab:RO-act-mVwe}
\end{figure}

\subsection{B\"{a}cklund auto-transformation}
Consider again the first and the second equations from the positive covering
of \Eref{eq:4} and replace $q_1$ by $v$ in them:
\begin{equation}
  v_t = \displaystyle{\frac{u_t}{u_y}},
  \qquad
  v_x = \displaystyle{\frac{1}{u_y}}.
  \label{BT_mVwe_1}
\end{equation}
This gives the following expressions for $u_t$ and $u_y$:
\begin{equation}
  u_t = \displaystyle{\frac{v_t}{v_x}},
  \qquad
  u_y = \displaystyle{\frac{1}{v_x}}.
  \label{BT_mVwe_2}
\end{equation}
Cross-differentiation of this system with respect to $y$ and $t$ gives
$v_{tx} = v_t v_{xy} - v_xv_{ty}$.  This equation differs from \Eref{eq:4}
just by the change of variables
\begin{equation}
  x \mapsto y, \quad y \mapsto x.
  \label{BT_mVwe_3}
\end{equation}
Thus we have
\begin{proposition}
  \label{thm:BT_mVwe}
  The superposition of~\eref{BT_mVwe_1} and~\eref{BT_mVwe_3} defines a
  B\"{a}cklund auto-trans\-for\-ma\-ti\-on for \Eref{eq:4}. The inverse transformation
  is given by the su\-per\-po\-si\-ti\-on of~\eref{BT_mVwe_3} and~\eref{BT_mVwe_2}.
\end{proposition}

\section{Conclusions}
\label{sec:discussion}

The equations discussed above have many common features:
\begin{enumerate}
\item all of them admit a differential coverings with non-removable parameter;
\item  all of them are linearly degenerate;
\item each of these equation can be obtained as a symmetry reduction of the 5D
  equation $u_{zs}+u_{yz}-u_{ts} + u_zu_{xs} - u_s u_{xz} = 0$,
  see~\cite{BKMV-Lob-2015};
\item as it is shown in~\cite{MorozovPavlov2016}, they are pair-wise related
  by B\"{a}cklund transformations.
\end{enumerate}
\begin{table}[t]
\caption{\label{tab:all-the-algebras}Lie algebras of nonlocal symmetries.}
\begin{indented}
\item[]\begin{tabular}{@{}l|c|c}
\br
    &$\tau^+$&$\tau^-$\\\hline
    \rule{0em}{12pt}rdDym equation
    &$\mathfrak{W}_0^-\ltimes(\mathfrak{L}_3^-[t]\oplus\mathfrak{L}[y])$
             &$\mathfrak{W}_0^+\ltimes\mathfrak{L}[t]\oplus\mathfrak{V}[y]$\\\hline
    \rule{0em}{12pt}3D Pavlov equation
    &$\mathfrak{W}_0^-\ltimes(\mathfrak{L}[q_0]\oplus\mathfrak{L}_4^-[t])$
             &$\mathfrak{W}_{-1}^+\ltimes \mathfrak{L}[t]$\\\hline
    \rule{0em}{12pt}UHE
    &$\mathfrak{W}_0^+\ltimes(\mathfrak{L}_2^+[x]\oplus\mathfrak{L}[t])$
             &$\mathfrak{W}_0^-\ltimes(\mathfrak{L}_2^-[t]\oplus\mathfrak{L}[x])$\\\hline
    \rule{0em}{12pt}mVwe
    &$\widetilde{\mathfrak{W}}_0^+
      \ltimes(\mathfrak{L}[y]\oplus\mathfrak{L}_2^+[x])
      \oplus \mathfrak{V}[t]$
             &$\widetilde{\mathfrak{W}}_0^- \ltimes\mathfrak{L}[x]\oplus \mathfrak{V}[y]
               \oplus \mathfrak{V}[t]$\\
\br
\end{tabular}
\end{indented}
\end{table}
This similarity manifests itself in striking resemblance of their symmetry
algebra structures (see \Tref{tab:all-the-algebras}). Perhaps, the mVWE
equation stands alone a bit: its symmetries are not graded in the same sense
as symmetries of the other three equations.

We think that it will be extremely interesting to find out which properties of
equations, besides their linearly degeneracy, are responsible for the such
symmetry structures and plan to shed light on this problem in future
research. We also intend to clarify the invariant meaning of the operators
$\mathcal{Y}$ that played such an important role in the above discussed
constructions.

\ack

Computations were done using the \textsc{Jets} software,~\cite{Jets}. The
second author (ISK) was partially supported by the Simons-IUM fellowship. OIM
is greatly indebted to  for a
financial support. HB and PV were supported by RVO funding for I\v{C}47813059.


\section*{References}

\begin{thebibliography}{99}
\bibitem{A-Sh} Adler V E and Shabat A B 2007 \emph{Theor.\ Math.\ Phys.}
  \textbf{153} 1373--87.

\bibitem{BKMV-Lob-2015} Baran H, Krasil{\cprime}shchik I S, Morozov O I, and
  Voj\v{c}\'{a}k P 2015  \emph{Lobachevskii J.\ of Math.}
  \textbf{36} 225--33

\bibitem{BKMV-2014} Baran H, Krasil{\cprime}shchik I S, Morozov O I, and
  Voj\v{c}\'{a}k P 2014 \emph{J.\ of Nonlinear Math.\ Phys.} \textbf{21}
  643--71 (arXiv:1407.0246 [nlin.SI])

\bibitem{BKMV-2015} Baran H, Krasil{\cprime}shchik I S, Morozov O I, and
  Voj\v{c}\'{a}k P 2015 \emph{J.\ of Nonlinear Math.\ Phys.}
  \textbf{22} 210--32 (arXiv:1412.6461 [nlin.SI])

\bibitem{BKMV-2016} Baran H, Krasil{\cprime}shchik I S, Morozov O I, and
  Voj\v{c}\'{a}k P 2016 \emph{Theor.\ Math.\ Phys.} \textbf{188} 1273--95
  (arXiv:1507.00897 [nlin.SI])

\bibitem{Jets} Baran H and Marvan M Jets. A software for differential calculus
  on jet spaces and diffieties http://jets.math.slu.cz

\bibitem{Blaszak} B{\l}aszak M 2002 \emph{Phys.\ Lett.\ A} \textbf{297}
  191--5.

\bibitem{AMS} Bocharov A V \etal 1997 \emph{Symmetries of Differential
    Equations in Mathematical Physics and Natural Sciences}, edited by
  A.M.~Vinogradov and I.S.~Krasil{\cprime}shchik) (Moscow: Factorial Publ.\
  House, in Russian). English translation: Amer.\ Math.\ Soc., 1999

\bibitem{Dun} Dunajski M 2004 \emph{J.\ Geom.\ Phys.} \textbf{51} 126--37

\bibitem{V-Web-1} Dunajski M and Kry\'{n}ski W 2014 \emph{Math.\ Proc.\ of the
    Cambridge Philosophical Soc.}, \textbf{157} 139--50

\bibitem{Fer-Mos} Ferapontov E V and Moss J 2015 \emph{Commun.\ in Analysis
    and Geometry} \textbf{23} 91--127 (arXiv:1204.2777 [math.DG])

\bibitem{GT} Gibbons J annd Tsarev S P 1996 \emph{Phys.\ Lett.\ A}
  \textbf{211} 19--24

\bibitem{VinKrasTrends} Krasil{\cprime}shchik I S and Vinogradov A M 1989
  \emph{Acta Appl.\ Math.} \textbf{15} 161--209

\bibitem{KruglikovMorozov2015} Kruglikov B and Morozov O 2015 \emph{Lett.\
    Math.\ Phys.} \textbf{105} 1703--23

\bibitem{MalykhNutkuSheftel2004} Malykh A A, Nutku Y, and Sheftel M B 2004
  \emph{J.\ Phys.\ A} \textbf{37} 7527--46

\bibitem{MartinezAlonsoShabat2002} Mart{\'{\i}}nez Alonso L and Shabat A B
  2002 \emph{Phys.\ Lett.\ A} \textbf{299} 359--65

\bibitem{MartinezAlonsoShabat2004} Mart{\'{\i}}nez Alonso L and Shabat A B
  2004 \emph{Theor.\ Math.\ Phys.}  \textbf{140} 1073--85

\bibitem{Mar-another} Marvan M 1996 \emph{Differential Geometry and
    Applications, Proc.\ Conf.\ Brno 1995} (Masaryk University Brno 1996)
  pp.~393--402

\bibitem{MarvanSergyeyev2012} Marvan M amd Sergyeyev A 2012 \emph{Inverse
    Probl.} \textbf{28} 025011

\bibitem{Morozov2012} Morozov O I 2012 Recursion operators and nonlocal
  symmetries for integrable rmdKP and rdDym equations arXiv:1202.2308
  [nlin.SI]

\bibitem{Morozov2014} Morozov O I 2014 \emph{Cent. Eur. J. Math.} \textbf{12}
  271--83

\bibitem{MorozovPavlov2016} Morozov O I and Pavlov M V 2016 B\"{a}cklund
  transformations between four Lax-integrable 3D equations arXiv:1611.04036
  [nlin.SI]

\bibitem{MorozovSergyeyev2014} Morozov O I and Sergyeyev A 2014
  \emph{J. Geom. Phys.} \textbf{85} 40--5

\bibitem{Ovsienko2010} Ovsienko V 2010 \emph{Adv.\ Pure Appl.\ Math.}
  \textbf{1} 7--17.

\bibitem{Pavlov2006} Pavlov M V 2006 \emph{Intern.\ Math.\ Research Notes}
  \textbf{2006} article ID 46987 1--43.

\bibitem{Pav} Pavlov M V 2003 \emph{J.\ Math.\ Phys.} \textbf{44} 4134--56.

\bibitem{Schlichenmaier2014} Schlichenmaier M 2014 \emph{Krichever--Novikov
    type algebras. Theory and applications} (Berlin/Boston: De Gruyter)

\bibitem{Sergyeyev2015} Sergyeyev A 2015 A Simple Construction of Recursion
  Operators for Multidimensional Dispersionless Integrable Systems
  arXiv:1501.01955

\bibitem{V-Web-2} Zakharevich I 2000 Nonlinear wave equation, nonlinear
  Riemann problem, and the twistor transform of Veronese webs
  arXiv:math-ph/0006001
\end{thebibliography}
\end{document}